\documentclass[10pt,twocolumn,letterpaper]{article}
\usepackage[letterpaper,top=0.75in,bottom=1.25in,left=0.75in,right=0.75in,columnsep=0.33in]{geometry}
\usepackage{times}
\usepackage{helvet}
\usepackage{courier}
\usepackage[hyphens]{url}
\usepackage{graphicx}
\usepackage{amsmath}
\usepackage{amssymb}
\usepackage{array}
\usepackage{caption}
\usepackage{float}
\usepackage{tcolorbox}
\usepackage{xcolor}
\usepackage[numbers,compress]{natbib}
\usepackage[hidelinks]{hyperref}
\newcommand{\ph}[1]{\textcolor{red}{[#1]}}
\newcommand{\AUCBlock}{\ph{AUCBlock}}
\newcommand{\AUCLoso}{\ph{AUCLoso}}
\newcommand{\AUCNight}{\ph{AUCNight}}
\newcommand{\AUCOldBlock}{\ph{AUCOldBlock}}
\newcommand{\AUCRandom}{\ph{AUCRandom}}

\newcommand{\AggVerdict}{\ph{AggVerdict}}
\newcommand{\BiasDirect}{\ph{BiasDirect}}
\newcommand{\BiasKAll}{\ph{BiasKAll}}
\newcommand{\BiasKOrig}{\ph{BiasKOrig}}
\newcommand{\BiasKPts}{\ph{BiasKPts}}
\newcommand{\BrierBlock}{\ph{BrierBlock}}

\newcommand{\CalSlope}{\ph{CalSlope}}

\newcommand{\CorrAge}{\ph{CorrAge}}
\newcommand{\CorrDist}{\ph{CorrDist}}
\newcommand{\CorrRain}{\ph{CorrRain}}
\newcommand{\ECEBlock}{\ph{ECEBlock}}

\newcommand{\GapMAEKrig}{\ph{GapMAEKrig}}
\newcommand{\GapMAEState}{\ph{GapMAEState}}
\newcommand{\GapVerdict}{\ph{GapVerdict}}
\newcommand{\NConflict}{\ph{NConflict}}
\newcommand{\NMapped}{\ph{NMapped}}
\newcommand{\NUnknown}{\ph{NUnknown}}
\newcommand{\NLate}{\ph{NLate}}
\newcommand{\NLateUnknown}{\ph{NLateUnknown}}
\newcommand{\MeanPUnknown}{\ph{MeanPUnknown}}
\newcommand{\RhoUnkSens}{\ph{RhoUnkSens}}
\newcommand{\OvUnk}{\ph{OvUnk}}
\newcommand{\NBottomLate}{\ph{NBottomLate}}
\newcommand{\IntCovHeldLGA}{\ph{IntCovHeldLGA}}
\newcommand{\IntCovHeldState}{\ph{IntCovHeldState}}
\newcommand{\IntCovHeldLGATen}{\ph{IntCovHeldLGATen}}
\newcommand{\IntCovHeldStateTen}{\ph{IntCovHeldStateTen}}
\newcommand{\NIntHeld}{\ph{NIntHeld}}
\newcommand{\NRound}{\ph{NRound}}
\newcommand{\PctRound}{\ph{PctRound}}
\newcommand{\RoundPctFunc}{\ph{RoundPctFunc}}
\newcommand{\RoundPctNotInUse}{\ph{RoundPctNotInUse}}
\newcommand{\AUCRoundBlock}{\ph{AUCRoundBlock}}
\newcommand{\AUCRoundLoso}{\ph{AUCRoundLoso}}
\newcommand{\BrierRound}{\ph{BrierRound}}
\newcommand{\ECERound}{\ph{ECERound}}
\newcommand{\CalSlopeRound}{\ph{CalSlopeRound}}
\newcommand{\NRoundRanked}{\ph{NRoundRanked}}
\newcommand{\NRoundCommon}{\ph{NRoundCommon}}
\newcommand{\RhoRound}{\ph{RhoRound}}
\newcommand{\OvRound}{\ph{OvRound}}
\newcommand{\RhoRoundAlt}{\ph{RhoRoundAlt}}
\newcommand{\OvRoundAlt}{\ph{OvRoundAlt}}
\newcommand{\LosoMax}{\ph{LosoMax}}
\newcommand{\LosoMaxState}{\ph{LosoMaxState}}
\newcommand{\LosoMin}{\ph{LosoMin}}
\newcommand{\LosoMinState}{\ph{LosoMinState}}
\newcommand{\LosoWeakStates}{\ph{LosoWeakStates}}
\newcommand{\MAECountDirect}{\ph{MAECountDirect}}
\newcommand{\MAECountKAll}{\ph{MAECountKAll}}
\newcommand{\MAECountKOrig}{\ph{MAECountKOrig}}
\newcommand{\MAECountKPts}{\ph{MAECountKPts}}
\newcommand{\MAEDirect}{\ph{MAEDirect}}
\newcommand{\MAEKAll}{\ph{MAEKAll}}
\newcommand{\MAEKOrig}{\ph{MAEKOrig}}
\newcommand{\MAEKPts}{\ph{MAEKPts}}
\newcommand{\MedAgeObs}{\ph{MedAgeObs}}
\newcommand{\MedBeyondPct}{\ph{MedBeyondPct}}
\newcommand{\MedIntWidthPct}{\ph{MedIntWidthPct}}
\newcommand{\MedPtsPerK}{\ph{MedPtsPerK}}
\newcommand{\MedR}{\ph{MedR}}
\newcommand{\MedRChange}{\ph{MedRChange}}
\newcommand{\MedRFuture}{\ph{MedRFuture}}
\newcommand{\MedRpct}{\ph{MedRpct}}

\newcommand{\MicsN}{\ph{MicsN}}
\newcommand{\MicsPRain}{\ph{MicsPRain}}
\newcommand{\MicsPS}{\ph{MicsPS}}
\newcommand{\MicsRhoRain}{\ph{MicsRhoRain}}
\newcommand{\MicsRhoS}{\ph{MicsRhoS}}
\newcommand{\NAdequate}{\ph{NAdequate}}
\newcommand{\NBlocks}{\ph{NBlocks}}

\newcommand{\NFunc}{\ph{NFunc}}
\newcommand{\NInsufficient}{\ph{NInsufficient}}
\newcommand{\NLGAEval}{\ph{NLGAEval}}

\newcommand{\NLosoWeak}{\ph{NLosoWeak}}
\newcommand{\NNegAge}{\ph{NNegAge}}
\newcommand{\NNonFunc}{\ph{NNonFunc}}
\newcommand{\NNotInUse}{\ph{NNotInUse}}
\newcommand{\NPts}{\ph{NPts}}
\newcommand{\NRanked}{\ph{NRanked}}
\newcommand{\NRegistry}{\ph{NRegistry}}
\newcommand{\NScope}{\ph{NScope}}
\newcommand{\NSparse}{\ph{NSparse}}
\newcommand{\NTilesAll}{\ph{NTilesAll}}
\newcommand{\NTilesTwo}{\ph{NTilesTwo}}
\newcommand{\NVerify}{\ph{NVerify}}
\newcommand{\NVerifyFrame}{\ph{NVerifyFrame}}
\newcommand{\NVerifyLGAs}{\ph{NVerifyLGAs}}
\newcommand{\NVerifyStrata}{\ph{NVerifyStrata}}
\newcommand{\NotInUseShareChange}{\ph{NotInUseShareChange}}

\newcommand{\ZeroText}{0}
\newcommand{\OvAlt}{\ph{OvAlt}}
\newcommand{\OvKrig}{\ph{OvKrig}}
\newcommand{\OvObs}{\ph{OvObs}}
\newcommand{\OvRaw}{\ph{OvRaw}}
\newcommand{\OvRef}{\ph{OvRef}}
\newcommand{\OvThr}{\ph{OvThr}}
\newcommand{\PctAgeMissing}{\ph{PctAgeMissing}}
\newcommand{\PctBeyondNinety}{\ph{PctBeyondNinety}}

\newcommand{\PctRainAligned}{\ph{PctRainAligned}}
\newcommand{\PctRecentPts}{\ph{PctRecentPts}}
\newcommand{\PctSingleTiles}{\ph{PctSingleTiles}}
\newcommand{\PctStalePts}{\ph{PctStalePts}}
\newcommand{\PopCorr}{\ph{PopCorr}}
\newcommand{\RainSource}{\ph{RainSource}}
\newcommand{\RecentCut}{\ph{RecentCut}}
\newcommand{\RefShareChange}{\ph{RefShareChange}}
\newcommand{\RepYearMax}{\ph{RepYearMax}}
\newcommand{\RepYearMedian}{\ph{RepYearMedian}}
\newcommand{\RepYearMin}{\ph{RepYearMin}}
\newcommand{\RhoAltSens}{\ph{RhoAltSens}}
\newcommand{\RhoDirect}{\ph{RhoDirect}}
\newcommand{\RhoKAll}{\ph{RhoKAll}}
\newcommand{\RhoKOrig}{\ph{RhoKOrig}}
\newcommand{\RhoKPts}{\ph{RhoKPts}}
\newcommand{\RhoKrigSens}{\ph{RhoKrigSens}}
\newcommand{\RhoObsSens}{\ph{RhoObsSens}}
\newcommand{\RhoRawSens}{\ph{RhoRawSens}}
\newcommand{\RhoRefSens}{\ph{RhoRefSens}}
\newcommand{\RhoThrSens}{\ph{RhoThrSens}}
\newcommand{\SMax}{\ph{SMax}}
\newcommand{\SMean}{\ph{SMean}}
\newcommand{\SMin}{\ph{SMin}}
\newcommand{\SSD}{\ph{SSD}}
\newcommand{\STopFive}{\ph{STopFive}}
\newcommand{\STopOverlap}{\ph{STopOverlap}}

\newcommand{\TempCutoff}{\ph{TempCutoff}}
\newcommand{\TempMeanP}{\ph{TempMeanP}}
\newcommand{\TempNTest}{\ph{TempNTest}}

\newcommand{\TempYears}{\ph{TempYears}}
\newcommand{\TopStates}{\ph{TopStates}}
\newcommand{\VarioNugget}{\ph{VarioNugget}}
\newcommand{\VarioRange}{\ph{VarioRange}}
\newcommand{\VarioSill}{\ph{VarioSill}}
\newcommand{\VerifyMargin}{\ph{VerifyMargin}}
\newcommand{\TabBottomTen}{\multicolumn{7}{|c|}{\ph{rows written by revision\_analysis.py}} \\ \hline}
\newcommand{\TabBottomThirty}{\multicolumn{8}{|c|}{\ph{rows written by revision\_analysis.py}} \\ \hline}
\newcommand{\TabInsufficient}{\multicolumn{5}{|c|}{\ph{rows written by revision\_analysis.py}} \\ \hline}
\newcommand{\TabStateAUC}{\multicolumn{5}{|c|}{\ph{rows written by revision\_analysis.py}} \\ \hline}

\renewcommand{\AUCBlock}{0.747}
\renewcommand{\AUCLoso}{0.722}
\renewcommand{\AUCNight}{0.749}
\renewcommand{\AUCOldBlock}{0.747}
\renewcommand{\AUCRandom}{0.781}
\renewcommand{\AUCRoundBlock}{0.742}
\renewcommand{\AUCRoundLoso}{0.709}

\renewcommand{\AggVerdict}{the direct sum had the lowest error}
\renewcommand{\BiasDirect}{0.004}
\renewcommand{\BiasKAll}{0.002}
\renewcommand{\BiasKOrig}{$-$0.001}
\renewcommand{\BiasKPts}{0.003}
\renewcommand{\BrierBlock}{0.190}
\renewcommand{\BrierRound}{0.191}

\renewcommand{\CalSlope}{0.92}
\renewcommand{\CalSlopeRound}{0.92}

\renewcommand{\CorrAge}{$-$0.03}
\renewcommand{\CorrDist}{0.01}
\renewcommand{\CorrRain}{$-$0.19}
\renewcommand{\ECEBlock}{0.021}
\renewcommand{\ECERound}{0.022}

\renewcommand{\GapMAEKrig}{0.128}
\renewcommand{\GapMAEState}{0.123}
\renewcommand{\GapVerdict}{did not improve on}
\renewcommand{\IntCovHeldLGA}{89}
\renewcommand{\IntCovHeldLGATen}{89}
\renewcommand{\IntCovHeldState}{88}
\renewcommand{\IntCovHeldStateTen}{88}
\renewcommand{\LosoMax}{0.94}
\renewcommand{\LosoMaxState}{Borno}
\renewcommand{\LosoMin}{0.60}
\renewcommand{\LosoMinState}{Zamfara}
\renewcommand{\LosoWeakStates}{Zamfara, Kaduna, FCT, Bauchi, Yobe, Kano, Katsina, Gombe, Benue, Kebbi, Sokoto, Cross River, Niger, Jigawa, Ebonyi}
\renewcommand{\MAECountDirect}{13.7}
\renewcommand{\MAECountKAll}{14.1}
\renewcommand{\MAECountKOrig}{13.9}
\renewcommand{\MAECountKPts}{13.6}
\renewcommand{\MAEDirect}{0.118}
\renewcommand{\MAEKAll}{0.121}
\renewcommand{\MAEKOrig}{0.120}
\renewcommand{\MAEKPts}{0.119}
\renewcommand{\MeanPUnknown}{26.1}
\renewcommand{\MedAgeObs}{5}
\renewcommand{\MedBeyondPct}{91.5}
\renewcommand{\MedIntWidthPct}{41}
\renewcommand{\MedPtsPerK}{0.51}
\renewcommand{\MedR}{0.085}
\renewcommand{\MedRChange}{$-$9.6}
\renewcommand{\MedRFuture}{0.077}
\renewcommand{\MedRpct}{8.5}

\renewcommand{\MicsN}{37}
\renewcommand{\MicsPRain}{0.010}
\renewcommand{\MicsPS}{0.065}
\renewcommand{\MicsRhoRain}{0.418}
\renewcommand{\MicsRhoS}{0.307}
\renewcommand{\NAdequate}{580}
\renewcommand{\NBlocks}{94}

\renewcommand{\NBottomLate}{7}

\renewcommand{\NConflict}{761}

\renewcommand{\NFunc}{53,015}
\renewcommand{\NInsufficient}{12}
\renewcommand{\NIntHeld}{602}
\renewcommand{\NLGAEval}{575}
\renewcommand{\NLate}{13,434}
\renewcommand{\NLateUnknown}{12,560}

\renewcommand{\NLosoWeak}{15}
\renewcommand{\NMapped}{99,245}
\renewcommand{\NNegAge}{6}
\renewcommand{\NNonFunc}{33,670}
\renewcommand{\NNotInUse}{52,933}
\renewcommand{\NPts}{86,685}
\renewcommand{\NRanked}{580}
\renewcommand{\NRegistry}{34}
\renewcommand{\NRound}{85,643}
\renewcommand{\NRoundCommon}{568}
\renewcommand{\NRoundRanked}{568}
\renewcommand{\NScope}{614}
\renewcommand{\NSparse}{22}
\renewcommand{\NTilesAll}{22,952}
\renewcommand{\NTilesTwo}{14,049}
\renewcommand{\NUnknown}{12,560}
\renewcommand{\NVerify}{300}
\renewcommand{\NVerifyFrame}{2,175}
\renewcommand{\NVerifyLGAs}{60}
\renewcommand{\NVerifyStrata}{6}
\renewcommand{\NotInUseShareChange}{$-$99.9}

\renewcommand{\OvAlt}{12}
\renewcommand{\OvKrig}{50}
\renewcommand{\OvObs}{42}
\renewcommand{\OvRaw}{44}
\renewcommand{\OvRef}{53}
\renewcommand{\OvRound}{54}
\renewcommand{\OvRoundAlt}{9}
\renewcommand{\OvThr}{49}
\renewcommand{\OvUnk}{57}
\renewcommand{\PctAgeMissing}{52.2}
\renewcommand{\PctBeyondNinety}{58}

\renewcommand{\PctRainAligned}{100.0}
\renewcommand{\PctRecentPts}{0.1}
\renewcommand{\PctRound}{98.8}
\renewcommand{\PctSingleTiles}{38.8}
\renewcommand{\PctStalePts}{0.2}
\renewcommand{\PopCorr}{0.986}
\renewcommand{\RainSource}{annual CHIRPS totals for 2010 to 2024}
\renewcommand{\RecentCut}{2020}
\renewcommand{\RefShareChange}{$-$5.7}
\renewcommand{\RepYearMax}{2025}
\renewcommand{\RepYearMedian}{2015}
\renewcommand{\RepYearMin}{2010}
\renewcommand{\RhoAltSens}{0.332}
\renewcommand{\RhoDirect}{0.671}
\renewcommand{\RhoKAll}{0.674}
\renewcommand{\RhoKOrig}{0.675}
\renewcommand{\RhoKPts}{0.683}
\renewcommand{\RhoKrigSens}{0.990}
\renewcommand{\RhoObsSens}{0.954}
\renewcommand{\RhoRawSens}{0.939}
\renewcommand{\RhoRefSens}{0.982}
\renewcommand{\RhoRound}{0.989}
\renewcommand{\RhoRoundAlt}{0.255}
\renewcommand{\RhoThrSens}{0.964}
\renewcommand{\RhoUnkSens}{0.990}
\renewcommand{\RoundPctFunc}{61.8}
\renewcommand{\RoundPctNotInUse}{99.8}
\renewcommand{\SMax}{71.0}
\renewcommand{\SMean}{36.4}
\renewcommand{\SMin}{14.0}
\renewcommand{\SSD}{10.9}
\renewcommand{\STopFive}{Andoni (Rivers), Okpe (Delta), Etche (Rivers), Southern Ijaw (Bayelsa), Ethiope West (Delta)}
\renewcommand{\STopOverlap}{18}

\renewcommand{\TabBottomTen}{Mafa & Borno & 69,750 & 10 & 0.0025 & 0.0000 to 0.0226 & 2019 (6) \\ \hline
Gwoza & Borno & 310,772 & 123 & 0.0027 & 0.0000 to 0.0272 & 2019 (6) \\ \hline
Etche & Rivers & 590,817 & 19 & 0.0032 & 0.0007 to 0.0077 & 2015 (10) \\ \hline
Shelleng & Adamawa & 412,427 & 10 & 0.0035 & 0.0015 to 0.0069 & 2015 (10) \\ \hline
Bama & Borno & 346,942 & 136 & 0.0047 & 0.0000 to 0.0290 & 2019 (6) \\ \hline
Sapele & Delta & 388,082 & 12 & 0.0051 & 0.0027 to 0.0093 & 2015 (10) \\ \hline
Ahoada East & Rivers & 294,952 & 19 & 0.0067 & 0.0016 to 0.0158 & 2015 (10) \\ \hline
Yola South & Adamawa & 442,210 & 17 & 0.0071 & 0.0040 to 0.0115 & 2015 (10) \\ \hline
Okpe & Delta & 277,592 & 19 & 0.0071 & 0.0016 to 0.0167 & 2015 (10) \\ \hline
Ngala & Borno & 160,671 & 120 & 0.0074 & 0.0000 to 0.0537 & 2019 (6) \\ \hline}
\renewcommand{\TabBottomThirty}{1 & Mafa & Borno & 69,750 & 10 & 0.0025 & 0.0000 to 0.0226 & 2019 (6) \\ \hline
2 & Gwoza & Borno & 310,772 & 123 & 0.0027 & 0.0000 to 0.0272 & 2019 (6) \\ \hline
3 & Etche & Rivers & 590,817 & 19 & 0.0032 & 0.0007 to 0.0077 & 2015 (10) \\ \hline
4 & Shelleng & Adamawa & 412,427 & 10 & 0.0035 & 0.0015 to 0.0069 & 2015 (10) \\ \hline
5 & Bama & Borno & 346,942 & 136 & 0.0047 & 0.0000 to 0.0290 & 2019 (6) \\ \hline
6 & Sapele & Delta & 388,082 & 12 & 0.0051 & 0.0027 to 0.0093 & 2015 (10) \\ \hline
7 & Ahoada East & Rivers & 294,952 & 19 & 0.0067 & 0.0016 to 0.0158 & 2015 (10) \\ \hline
8 & Yola South & Adamawa & 442,210 & 17 & 0.0071 & 0.0040 to 0.0115 & 2015 (10) \\ \hline
9 & Okpe & Delta & 277,592 & 19 & 0.0071 & 0.0016 to 0.0167 & 2015 (10) \\ \hline
10 & Ngala & Borno & 160,671 & 120 & 0.0074 & 0.0000 to 0.0537 & 2019 (6) \\ \hline
11 & Ganye & Adamawa & 264,124 & 14 & 0.0082 & 0.0040 to 0.0156 & 2015 (10) \\ \hline
12 & Ethiope West & Delta & 332,156 & 25 & 0.0085 & 0.0025 to 0.0191 & 2015 (10) \\ \hline
13 & Ika South & Delta & 207,635 & 17 & 0.0087 & 0.0022 to 0.0202 & 2015 (10) \\ \hline
14 & Burutu & Delta & 146,235 & 12 & 0.0089 & 0.0024 to 0.0205 & 2015 (10) \\ \hline
15 & Ethiope East & Delta & 351,255 & 26 & 0.0091 & 0.0032 to 0.0195 & 2015 (10) \\ \hline
16 & Ahoada West & Rivers & 196,758 & 15 & 0.0092 & 0.0031 to 0.0199 & 2015 (10) \\ \hline
17 & Dikwa & Borno & 119,205 & 96 & 0.0097 & 0.0000 to 0.0757 & 2019 (6) \\ \hline
18 & Arochukwu & Abia & 293,641 & 29 & 0.0100 & 0.0022 to 0.0239 & 2015 (10) \\ \hline
19 & Ogba/Egbema/Ndoni & Rivers & 684,061 & 68 & 0.0101 & 0.0018 to 0.0182 & 2015 (10) \\ \hline
20 & Ohafia & Abia & 364,682 & 40 & 0.0101 & 0.0010 to 0.0191 & 2015 (10) \\ \hline
21 & Southern Ijaw & Bayelsa & 227,469 & 30 & 0.0102 & 0.0000 to 0.0210 & 2015 (10) \\ \hline
22 & Unuimo & Imo & 79,869 & 10 & 0.0103 & 0.0003 to 0.0279 & 2015 (10) \\ \hline
23 & Egbado South & Ogun & 495,805 & 30 & 0.0103 & 0.0053 to 0.0153 & 2015 (10) \\ \hline
24 & Ughelli South & Delta & 476,245 & 29 & 0.0104 & 0.0056 to 0.0183 & 2015 (10) \\ \hline
25 & Fufore & Adamawa & 319,905 & 14 & 0.0106 & 0.0071 to 0.0131 & 2015 (10) \\ \hline
26 & Bungudu & Zamfara & 568,987 & 35 & 0.0111 & 0.0060 to 0.0162 & 2015 (10) \\ \hline
27 & Ilaje & Ondo & 354,743 & 36 & 0.0112 & 0.0028 to 0.0195 & 2015 (10) \\ \hline
28 & Aninri & Enugu & 240,165 & 40 & 0.0114 & 0.0000 to 0.0250 & 2018 (7) \\ \hline
29 & Damboa & Borno & 203,856 & 50 & 0.0114 & 0.0000 to 0.0315 & 2019 (6) \\ \hline
30 & Aniocha South & Delta & 209,480 & 14 & 0.0119 & 0.0065 to 0.0200 & 2015 (10) \\ \hline}
\renewcommand{\TabInsufficient}{Madagali & Adamawa & 244,101 & 0 & none \\ \hline
Abadam & Borno & 145,932 & 0 & none \\ \hline
Gubio & Borno & 295,435 & 0 & none \\ \hline
Guzamala & Borno & 203,918 & 0 & none \\ \hline
Kaga & Borno & 134,350 & 0 & none \\ \hline
Kala/Balge & Borno & 52,167 & 0 & none \\ \hline
Kukawa & Borno & 210,731 & 0 & none \\ \hline
Marte & Borno & 133,159 & 0 & none \\ \hline
Mobbar & Borno & 321,637 & 0 & none \\ \hline
Nganzai & Borno & 186,465 & 0 & none \\ \hline
Geidam & Yobe & 291,288 & 0 & none \\ \hline
Gujba & Yobe & 188,760 & 0 & none \\ \hline
Mubi North & Adamawa & 178,299 & 1 & 2015 \\ \hline
Mubi South & Adamawa & 231,373 & 1 & 2015 \\ \hline
Numan & Adamawa & 137,581 & 1 & 2015 \\ \hline
Ika North East & Delta & 304,578 & 1 & 2015 \\ \hline
Ogu/Bolo & Rivers & 35,173 & 1 & 2015 \\ \hline
Lamurde & Adamawa & 115,640 & 2 & 2015 \\ \hline
Bonny & Rivers & 222,721 & 2 & 2015 \\ \hline
Gulani & Yobe & 195,818 & 2 & 2015 \\ \hline
Kwaya Kusar & Borno & 116,691 & 3 & 2015 \\ \hline
Tai & Rivers & 133,707 & 3 & 2015 \\ \hline
Chikun & Kaduna & 1,404,953 & 4 & 2015 \\ \hline
Toungo & Adamawa & 85,344 & 5 & 2015 \\ \hline
Aniocha North & Delta & 146,271 & 6 & 2015 \\ \hline
Emohua & Rivers & 302,037 & 6 & 2015 \\ \hline
Jada & Adamawa & 305,594 & 7 & 2015 \\ \hline
Konduga & Borno & 311,215 & 7 & 2019 \\ \hline
Bomadi & Delta & 84,299 & 7 & 2015 \\ \hline
Warri South West & Delta & 48,484 & 7 & 2015 \\ \hline
Hong & Adamawa & 157,436 & 8 & 2015 \\ \hline
Degema & Rivers & 64,455 & 8 & 2015 \\ \hline
Yunusari & Yobe & 243,622 & 8 & 2015 \\ \hline
Magumeri & Borno & 317,873 & 9 & 2015 \\ \hline}
\renewcommand{\TabStateAUC}{Zamfara & 1,582 & 0.726 & 0.605 & 0.599 \\ \hline
Kaduna & 4,991 & 0.611 & 0.589 & 0.608 \\ \hline
FCT & 751 & 0.502 & 0.650 & 0.622 \\ \hline
Bauchi & 3,330 & 0.761 & 0.651 & 0.635 \\ \hline
Yobe & 1,343 & 0.854 & 0.607 & 0.635 \\ \hline
Kano & 7,171 & 0.782 & 0.643 & 0.641 \\ \hline
Katsina & 5,554 & 0.719 & 0.662 & 0.644 \\ \hline
Gombe & 2,306 & 0.492 & 0.657 & 0.648 \\ \hline
Benue & 2,597 & 0.524 & 0.647 & 0.661 \\ \hline
Kebbi & 2,478 & 0.505 & 0.669 & 0.668 \\ \hline
Sokoto & 1,705 & 0.537 & 0.662 & 0.673 \\ \hline
Cross River & 2,647 & 0.447 & 0.674 & 0.679 \\ \hline
Niger & 2,784 & 0.621 & 0.683 & 0.681 \\ \hline
Jigawa & 9,822 & 0.839 & 0.660 & 0.684 \\ \hline
Ebonyi & 1,956 & 0.610 & 0.657 & 0.686 \\ \hline
Ondo & 2,752 & 0.398 & 0.713 & 0.701 \\ \hline
Kwara & 3,586 & 0.530 & 0.706 & 0.709 \\ \hline
Plateau & 2,933 & 0.541 & 0.710 & 0.709 \\ \hline
Abia & 949 & 0.350 & 0.691 & 0.724 \\ \hline
Adamawa & 288 & 0.708 & 0.738 & 0.739 \\ \hline
Taraba & 2,532 & 0.553 & 0.749 & 0.749 \\ \hline
Oyo & 3,273 & 0.577 & 0.746 & 0.754 \\ \hline
Nasarawa & 1,688 & 0.591 & 0.741 & 0.756 \\ \hline
Kogi & 1,455 & 0.393 & 0.758 & 0.758 \\ \hline
Anambra & 455 & 0.532 & 0.745 & 0.762 \\ \hline
Osun & 4,757 & 0.559 & 0.761 & 0.762 \\ \hline
Lagos & 1,375 & 0.668 & 0.722 & 0.764 \\ \hline
Imo & 1,007 & 0.479 & 0.759 & 0.770 \\ \hline
Ekiti & 1,473 & 0.511 & 0.776 & 0.771 \\ \hline
Bayelsa & 310 & 0.384 & 0.789 & 0.774 \\ \hline
Akwa Ibom & 1,744 & 0.343 & 0.787 & 0.782 \\ \hline
Edo & 971 & 0.461 & 0.786 & 0.783 \\ \hline
Ogun & 1,068 & 0.481 & 0.802 & 0.788 \\ \hline
Delta & 680 & 0.443 & 0.788 & 0.795 \\ \hline
Rivers & 651 & 0.730 & 0.832 & 0.818 \\ \hline
Enugu & 385 & 0.545 & 0.816 & 0.827 \\ \hline
Borno & 1,307 & 0.328 & 0.949 & 0.942 \\ \hline}
\renewcommand{\TempCutoff}{2015}
\renewcommand{\TempMeanP}{26.8}
\renewcommand{\TempNTest}{874}

\renewcommand{\TempYears}{2016 to 2025}
\renewcommand{\TopStates}{Delta (8), Borno (6), Rivers (4)}
\renewcommand{\VarioNugget}{0.008}
\renewcommand{\VarioRange}{20.0}
\renewcommand{\VarioSill}{0.097}
\renewcommand{\VerifyMargin}{5.7}

\title{\Large\textbf{Screening Communal Water Point Functionality and Nominal Capacity in Rural Nigeria:\\
A Geospatial Machine Learning Approach Using an Open Water Point Registry}}
\author{
  Rafiu Adekoya Badekale\\
  Hamoye Foundation\\
  \texttt{ade.badekale@hamoye.org}
}
\date{}

\begin{document}

\maketitle

\begin{abstract}
Access to safe drinking water remains uneven across Nigeria. The communal water points that serve most rural households fail often and are monitored rarely. Planning bodies need estimates for each Local Government Area (LGA), which household surveys rarely provide. We develop a reproducible screening geospatial pipeline that combines the Water Point Data Exchange (WPDx-Plus) registry with WorldPop population grids, GRID3 settlement extents and CHIRPS rainfall, in order to direct field inspection and registry improvement. An XGBoost classifier trained on \NPts{} mapped water points estimates the probability that each point was functional when its status was recorded, using point age and rainfall aligned to that date. Statuses were recorded between \RepYearMin{} and \RepYearMax{} (median \RepYearMedian), and \PctRecentPts\% were recorded in \RecentCut{} or later. Most records after 2015 have an unknown status that WPDx's cleaned label codes as non-functional, and we treat them as unknown. Under spatial block cross-validation the classifier reached an area under the receiver operating characteristic curve (AUC) of \AUCBlock, with a mean of \AUCLoso{} when each state was held out. Expected functional points per LGA are the sum of point probabilities, an approach that we tested against ordinary kriging (mean absolute error \MAEDirect{} against \MAEKOrig). We rank the \NRanked{} rural and peri-urban LGAs that meet a minimum record-count threshold of ten mapped points by a nominal capacity ratio, shown with an empirical residual range and the age of its records, and list \NRegistry{} LGAs below the threshold separately for verification. The median ratio was \MedR. A low ratio may reflect incomplete mapping, water sources outside the registry or a real shortage, and the analysis cannot distinguish these. Of the \NFunc{} points recorded as functional, \NNotInUse{} are labelled \textit{Functional, not in use}, all from one contributor. WPDx's own field records water as present at almost all of them, but whether they were used could not be confirmed, so the rankings depend on how the label is read: when these points are counted as unavailable, only \OvAlt{} of the 60 priority LGAs remain among the 60 lowest. Using only the dominant 2015 survey round, the spatial block AUC was \AUCRoundBlock. The outputs are screening indicators, not measures of access.

\end{abstract}

\section{1. Introduction}
Universal and equitable access to safe and affordable drinking water is a core target of the 2030 Agenda for Sustainable Development \citep{un2015}. Progress toward this target remains slow in sub-Saharan Africa, where rural households depend heavily on communal water points such as handpumps, boreholes and protected wells \citep{jmp2023}. In Nigeria, the 2021 Multiple Indicator Cluster Survey found that 71.5\% of household members used a basic drinking water service, with state values ranging from 29.1\% in Taraba to 97.9\% in Lagos \citep{mics2021}. These figures describe states. They do not tell a planner which of the 774 Local Government Areas (LGAs) within a state should be inspected first, or where the record of existing infrastructure is too thin to say.

A second problem concerns the condition of the infrastructure that does exist. A large literature on rural water supply shows that a substantial share of handpumps in sub-Saharan Africa are non-functional at any given time, commonly reported at between one quarter and one third \citep{foster2020}. Functionality depends on the age of the installation, the management arrangement, the hydrogeological setting and the distance to the community that relies on it \citep{foster2013,whaley2017}. Counting mapped water points therefore overstates supply, because a point that appears in an inventory may no longer deliver water.

Open water point registries offer a route to finer spatial resolution. The Water Point Data Exchange (WPDx-Plus) compiles point-level records contributed by governments, non-governmental organisations and development partners, including the location, technology, management type and the status of each point on the date it was last recorded \citep{wpdx2024}. A contributed registry is not a census, however. Its coverage reflects where data collection programmes have operated, its records were collected in different years, and it excludes whole categories of supply, such as privately owned household boreholes, which are the dominant source in many Nigerian cities \citep{healy2020,gronwall2020}.

This study sits within a program of work on age-structured population and cross-sector capacity in Nigeria for 2025 to 2030 \citep{hamoye2026aipita}, and it complements a national composite index of infrastructure stress built from open geospatial data \citep{openhamoye2026}. Our goal was to build a screening tool that directs inspection and mapping effort to the LGAs where the mapped communal water system is thinnest or least certain, and to test each step against held-out data. To this end, our main contributions are summarized as follows,

\begin{itemize}
    \item We build a date-aware functionality pipeline in which point age and rainfall are aligned to the date each status was recorded.
    \item We validate the pipeline by spatial block and leave-one-state-out cross-validation, and check the calibration of its probabilities at point and LGA level.
    \item We separate LGAs ranked for inspection from LGAs that need registry improvement, label every output by its record count and the age of its records, treat 2030 as a population-growth scenario with unchanged supply, and design a stratified sample for independent field verification.
\end{itemize}

\section{2. Related Work}
Our approach builds on the measurement of rural water access, the study of water point functionality, machine learning for infrastructure condition, geospatial population and remote-sensing covariates, spatial interpolation and small-area estimation, probability calibration and uncertainty, and composite indices for screening.

\textbf{Measuring Rural Water Access.} The WHO and UNICEF Joint Monitoring Programme (JMP) defines the standard service ladder for drinking water, from surface water through unimproved, limited and basic services to safely managed services \citep{jmp2023}. National and subnational estimates are derived almost entirely from household surveys such as MICS and the Demographic and Health Surveys. These surveys measure what households report about their own supply, which makes them the reference standard for access, but their sampling designs rarely support estimates below the state or regional level \citep{mics2021}. The Local Burden of Disease WaSH collaborators mapped access to drinking water and sanitation at 5 km resolution by combining survey clusters with spatial covariates \citep{lbd2020}. This approach remains anchored to household survey data and does not describe the condition of the infrastructure that serves households.

\textbf{Water Point Functionality and Sustainability.} A parallel literature studies the infrastructure itself. Foster \citep{foster2013} analysed community-managed handpumps in Liberia, Sierra Leone and Uganda and found that functionality was associated with the age of the point, the presence of fee collection, the distance to spare parts and the governance arrangement. A later review of handpump data from sub-Saharan Africa and the Asia-Pacific region reported persistent non-functionality and large differences between countries and programmes \citep{foster2020}. Whaley and Cleaver \citep{whaley2017} cautioned that functionality, while measurable, captures only part of what makes rural supply sustainable. These studies motivate the use of functionality, not the mere presence of a point, as the supply-side quantity, and they also show that functionality changes over time, so a recorded status has a date that matters.

\textbf{Machine Learning for Water Point Condition.} Wilson et al. \citep{wilson2017} used ensemble models with sensor data from handpumps in Kenya to forecast failures and support preventive maintenance. Gradient-boosted decision trees, and XGBoost in particular \citep{chen2016xgboost}, are widely used for tabular prediction because they capture non-linear effects and interactions and handle missing values natively. A known risk in spatial applications is that random cross-validation overstates performance when nearby observations are correlated. Roberts et al. \citep{roberts2017} recommended blocked cross-validation for structured data, including data with temporal structure, and Ploton et al. \citep{ploton2020} showed that spatially independent validation can reveal much weaker performance than random validation in large-scale mapping models.

\textbf{Geospatial Population and Remote-Sensing Covariates.} WorldPop produces constrained, age- and sex-disaggregated population grids reconciled to national totals \citep{tatem2017,bondarenko2025}, and Leasure et al. \citep{leasure2020} developed a hierarchical Bayesian framework for national population mapping from sparse survey data, with Nigeria as a key application. CHIRPS combines satellite and station data into a rainfall record that starts in 1981 \citep{funk2015}, which allows rainfall to be matched to the year of an observation. NASA's Black Marble suite provides corrected nighttime light radiance \citep{roman2018}, a common proxy for economic activity \citep{henderson2012} that is contaminated by gas flaring in oil-producing regions \citep{elvidge2009}. Jean et al. \citep{jean2016} showed that dense satellite proxies combined with machine learning can predict outcomes where ground data are sparse.

\textbf{Spatial Interpolation and Small-Area Estimation.} Ordinary kriging estimates a value at an unsampled location as a weighted average of nearby observations, with weights derived from a fitted variogram \citep{oliver1990}. Small-area estimation methods combine survey benchmarks with auxiliary data for areas with little or no direct sampling \citep{rao2015}. Both rest on assumptions, spatial stationarity for kriging and a stable link between auxiliary data and the target for small-area estimation, that should be tested against held-out data rather than assumed.

\textbf{Probability Calibration and Uncertainty.} An estimate built by summing predicted probabilities is only as good as the calibration of those probabilities, meaning that events predicted at 70\% should occur about 70\% of the time. Niculescu-Mizil and Caruana \citep{niculescu2005} showed that boosted trees can produce distorted probabilities, and the Brier score \citep{brier1950} remains the standard summary of probabilistic accuracy. Conformal prediction builds distribution-free intervals from held-out residuals \citep{angelopoulos2023}. We borrow its residual-quantile construction to give each LGA an empirical residual range, but because our residuals come from the same LGAs, the ranges do not carry its coverage guarantee.

\textbf{Composite Indices for Screening.} Composite indicators combine normalised components into a single score and are widely used for screening and ranking \citep{oecd2008}. Cattaneo et al. \citep{cattaneo2021} mapped urban and rural catchment areas globally to reveal unequal access to services, and Weiss et al. \citep{weiss2018} mapped travel time to cities as a measure of accessibility. The OpenHamoye infrastructure stress index fused population, nightlights, building footprints and facility registries for kilometre-scale grid cells in Nigeria, and framed its outputs as proxy-based screening tools that do not measure engineered capacity \citep{openhamoye2026}. We adopt the same framing.

In this work, we combine the infrastructure-inventory tradition with explicit tests at every step: spatial and state-level validation of the classifier, a calibration check of its probabilities, a held-out comparison of aggregation methods, and empirical residual ranges for the final LGA values. To the best of our knowledge, no published study has produced a date-aware, validated screening of communal water point functionality for all of Nigeria's rural LGAs together with an explicit account of where the registry cannot support a ranking.

\section{3. Methodology}
\begin{figure}[!htbp]
    \centering
    \includegraphics[width=\linewidth]{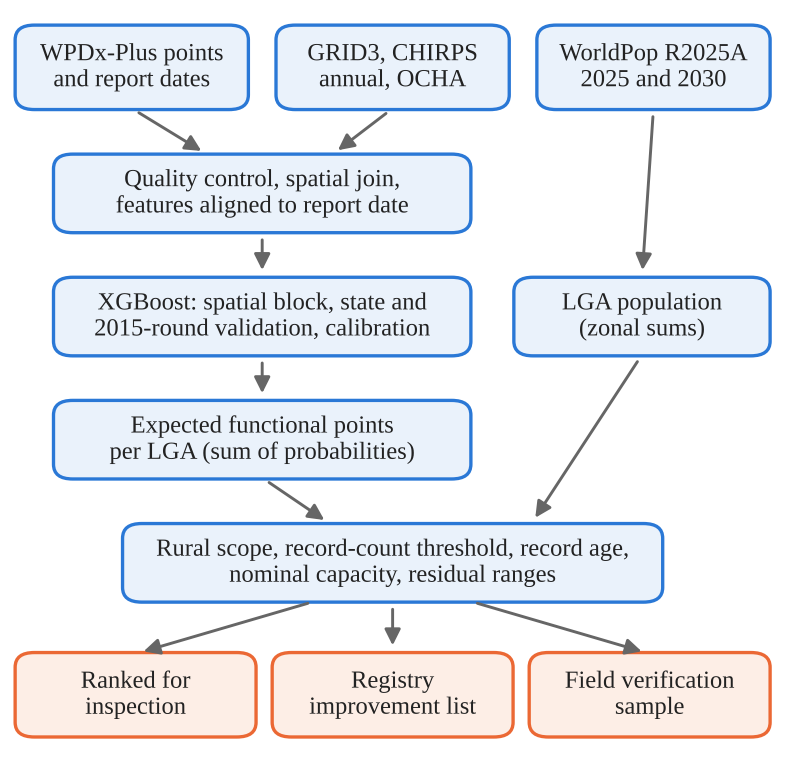}
    \caption{Flow diagram of the study.}
    \label{fig:flow}
\end{figure}

The study followed eight stages: data collection; dating of observations and alignment of features to those dates; functionality prediction; validation; aggregation to LGAs; nominal capacity, record-count threshold and record age; residual ranges and sensitivity analysis; and the design of a field verification sample. Figure~\ref{fig:flow} summarises the workflow.

\subsection{Datasets}
All inputs are open or freely available (Table~\ref{tab:data}). WPDx-Plus provides the location, water source type, technology, management arrangement, installation and rehabilitation years, reported status and the date of the status report for each mapped point in Nigeria \citep{wpdx2024}. WorldPop R2025A provides constrained age-sex population counts at 30 arc-seconds (about 1 km) for 2025 and 2030 \citep{bondarenko2025}. GRID3 Settlement Extents v4.1 provides 2,546,560 settlement blocks \citep{grid3settlement}. Administrative boundaries for the 774 LGAs come from the OCHA Common Operational Dataset \citep{ochacodab}. Rainfall comes from CHIRPS \citep{funk2015}, using \RainSource. Nightlight radiance from VIIRS Black Marble VNP46A3 \citep{roman2018} is used only in a sensitivity test. State-level basic drinking water service rates from the Nigeria MICS 2021 \citep{mics2021} are reported as context.

\begin{table}[!htbp]
    \centering
    \renewcommand{\arraystretch}{1.3}
    \resizebox{\columnwidth}{!}{%
    \begin{tabular}{|l|l|l|}
        \hline
        \textbf{Dataset} & \textbf{Resolution} & \textbf{Role} \\ \hline
        WPDx-Plus Nigeria & Point, dated & Status labels, report dates, features \\ \hline
        WorldPop R2025A age-sex & 1 km & Population, 2025 and 2030 \\ \hline
        GRID3 Settlement Extents v4.1 & Block polygons & Distance feature, scope check \\ \hline
        OCHA COD-AB boundaries & LGA polygons & Spatial unit of analysis \\ \hline
        CHIRPS v2.0 annual & 0.05$^\circ$ & Rainfall, aligned to report year \\ \hline
        VIIRS Black Marble VNP46A3 & 15 arc-sec & Sensitivity test only \\ \hline
        MICS 2021 (Table WS.1.2) & State & Context only \\ \hline
    \end{tabular}
    }
    \caption{Datasets used in the study.}
    \label{tab:data}
\end{table}

\subsection{Quality Control}
We checked every dataset for encoding artefacts and sentinel values. CHIRPS uses $-9999$, WorldPop $-99999$ and VIIRS $-999.9$ as no-data values, and each was masked before any arithmetic. We parsed the WPDx report date and treated report years outside 1990 to 2026 and installation years after 2026 as missing. For \NNegAge{} records the installation or rehabilitation year was later than the report year, so age at observation was set to missing. All joins between tables were keyed on the official LGA code (\texttt{adm2\_pcode}), because six LGA names (Bassa, Ifelodun, Irepodun, Nasarawa, Obi and Surulere) each occur in two states.

\subsection{Data Statistics}
The dataset contains \NMapped{} mapped water points, of which \NPts{} have a known status. We mapped WPDx's \texttt{status\_clean} field to a binary target, treating \textit{Functional}, \textit{Functional, needs repair} and \textit{Functional, not in use} as functional, and \textit{Non-Functional}, \textit{Non-Functional, dry season} and \textit{Abandoned/Decommissioned} as non-functional. This produced \NFunc{} functional and \NNonFunc{} non-functional points. Of the functional points, \NNotInUse{} (99.8\%) are recorded as \textit{Functional, not in use}, almost all of them in 2015, and only 82 as \textit{Functional} or \textit{Functional, needs repair}. All 52,933 come from one contributor, the Federal Ministry of Water Resources (FMWR), and they make up at least 99.8\% of the functional points it recorded. WPDx's own status field, which records whether water was present when a point was assessed, is \textit{Yes} for 52,874 of them (99.9\%), while the 82 points labelled \textit{Functional} or \textit{Functional, needs repair} all have \textit{No}. A label applied to almost every working point of one contributor, at points where water was present, reads more like that contributor's coding convention than a finding that the points were unused. The WPDx export does not include the contributor's original status text, so this cannot be confirmed. We count these points as functional in the main analysis and treat them as unavailable in a sensitivity test, because a point that works but is not used may not supply anyone. For \NConflict{} records the status field and the cleaned label disagree; they keep the cleaned label. All records carry the facility type \textit{Improved}, so the registry contains no unimproved sources or surface water.

\begin{figure}[!htbp]
    \centering
    \includegraphics[width=\linewidth]{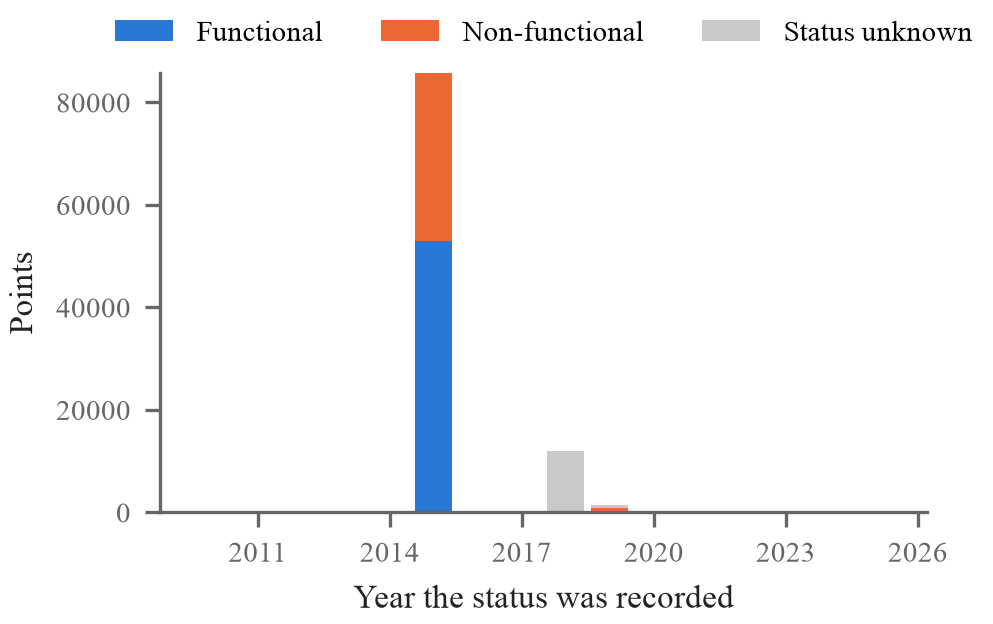}
    \caption{Year in which each water point status was recorded, by status. Status unknown: WPDx's own status field is \textit{Unknown}, although its cleaned label codes these points as non-functional.}
    \label{fig:years}
\end{figure}

\textbf{Observation dates.} Every point carries a usable report date. Statuses were recorded between \RepYearMin{} and \RepYearMax{}, with a median of \RepYearMedian, but they are concentrated in a few years (Figure~\ref{fig:years}). About 86\% were recorded in 2015 and \PctStalePts\% before it. The remaining \NLate{} records (13.5\%) date from 2016 onward, almost all before \RecentCut{}, since only \PctRecentPts\% date from \RecentCut{} or later. WPDx's cleaned label codes every one of them as non-functional. However, \NLateUnknown{} of them, contributed by GRID3 and reported in 2018 and 2019, have the value \textit{Unknown} in WPDx's own status field, so their non-functional label records a missing status, not an observed failure. We treat all \NUnknown{} points with an unknown status as mapped points whose status was not observed: they are excluded from training and validation, and they receive a predicted probability like any other point (mean \MeanPUnknown\%). The other \TempNTest{} statuses recorded after 2015, from iMMAP, WaterAid and other contributors, are all non-functional. A recorded status is therefore a historical observation: it describes the point when it was visited, not its condition today. Every functionality estimate in this paper is an estimate of status at the time of the last report.

\subsection{Data Preprocessing}
Each water point was assigned to an LGA by a point-in-polygon spatial join against the OCHA boundaries in WGS84. Thirty points fell just outside every LGA polygon, mostly along the borders with Cameroon and Niger, and were assigned using WPDx's own state and LGA fields. LGAs without any mapped point, mostly in Borno, Yobe and Adamawa, were labelled as having insufficient data. No supply was imputed for them. Bakassi, a legacy boundary entry for territory transferred to Cameroon after the 2002 judgment of the International Court of Justice \citep{icj2002}, was excluded.

\subsection{Date-Aligned Feature Engineering}
Each feature is measured, where the data allow, at the time the status was recorded (Appendix A).

\textit{Age at observation} is the report year minus the later of the installation and rehabilitation years. Values are missing for \PctAgeMissing\% of points. They were left missing, since XGBoost learns a default direction for missing values \citep{chen2016xgboost}.

\textit{Rainfall} enters twice: as the long-term mean annual total and as the total in the year the status was recorded, which was available for \PctRainAligned\% of dated points.

\textit{Distance to settlement} is the distance in metres to the boundary of the nearest GRID3 settlement block, computed in UTM zone 32N, with a binary \textit{inside-settlement} flag. The settlement layer describes the recent settlement pattern and cannot be dated to each observation. Settlement footprints change slowly compared with the reporting period, but this feature is the least well aligned.

\textit{Management type} is a one-hot encoding of WPDx's management categories.

The nightlight trend from 2018 to 2023 cannot be aligned with statuses recorded before 2018, so it is not a model feature. We report its effect as a sensitivity test.

\subsection{Functionality Model}
We trained an XGBoost classifier \citep{chen2016xgboost} with 200 trees, a maximum depth of 5 and a learning rate of 0.1, using scikit-learn for data handling and validation \citep{pedregosa2011}. The model estimates the probability that a point was functional on the date its status was recorded. Historical statuses cannot validate present-day functionality, so we make no claim about the condition of any point today. For a sensitivity test only, we also predicted every point with its age set to 2025 and its rainfall set to the long-term mean, to show how much the estimates depend on ageing. This is an assumption, not a forecast.

\subsection{Validation}
Validation proceeds from the most to the least demanding test, and spatial validation comes first.
\begin{enumerate}
    \item \textit{Spatial block cross-validation.} Points were grouped into \NBlocks{} cells of 1$^\circ$ by 1$^\circ$, and five folds were formed by cell, so no point was predicted by a model that had seen its neighbourhood \citep{roberts2017,ploton2020}.
    \item \textit{Leave-one-state-out validation.} The model was trained on all other states and tested on the held-out state, for every state with enough points.
    \item \textit{Random stratified cross-validation}, reported for reference only.
\end{enumerate}
For each scheme we report the AUC. For spatial block cross-validation we also check calibration with a reliability curve, the Brier score \citep{brier1950}, the expected calibration error (ECE) over ten probability bins, and the calibration slope, whose ideal value is 1 \citep{niculescu2005}. A temporal hold-out was also run, training on statuses recorded up to \TempCutoff{} and applying the model to the \TempNTest{} statuses recorded from \TempYears. Because all of these later statuses are non-functional, the AUC is undefined for this split, so it is reported only as a check of how the model rates later records, not as a validation. Because about 86\% of statuses come from one survey round, recorded in 2015, we also repeated spatial block and leave-one-state-out validation using only that round, and rebuilt the LGA ranking from its points alone (LGAs with at least ten 2015 points). This removes every post-2015 record and every difference between rounds. At LGA level, we compare the mean predicted probability with the observed share of functional points. The probabilities used in all LGA calculations are the out-of-fold predictions from spatial block cross-validation, so no point's probability comes from a model trained on its own status or on its neighbours.

\subsection{Aggregation to LGAs}
The expected number of functional points in LGA $i$ is
\begin{equation}
F_i = \sum_{j \in i} p_j ,
\end{equation}
where $p_j$ is the predicted probability for mapped point $j$. This requires no classification threshold and no interpolation. Points with an unknown status take their probability from the model fitted to all points of known status, and enter $F_i$ like any other mapped point. We compared it with an interpolation-based alternative, $F_i = P_i \times d_i$, where $P_i$ is the number of mapped points and $d_i$ the ordinary kriging prediction at the LGA centroid, fitted to 0.02$^\circ$ tile means after dropping tiles with a single point. These choices keep kriging computationally feasible and give values for LGAs without points, but they rest on assumptions that should be tested. For the \NLGAEval{} in-scope LGAs with at least ten points, we compared the observed share of functional points with four estimates: the direct mean of point probabilities; kriging from tiles with two or more points, evaluated at the LGA centroid; kriging from all tiles; and kriging evaluated at the LGA's own point locations. We also tested kriging in its gap-filling role by holding out whole LGAs in five folds and predicting each from the remaining tiles, against a baseline of the mean observed share in other LGAs of the same state. Kriging would be retained only if it reduced error.

\subsection{Nominal Capacity and the 2030 Scenario}
The nominal capacity of LGA $i$ and its ratio to population are
\begin{equation}
C_i = k\,F_i, \qquad R_i = \frac{C_i}{\text{Pop}_{i,2025}}, \qquad B_i = 1 - \min(R_i, 1),
\end{equation}
where $k = 300$ people per point is a planning value within the Sphere range of 250 for a tap and 500 for a handpump \citep{sphere2018}, and $B_i$ is the share of the population beyond the nominal capacity of mapped functional points. The value $k$ converts a count of points into the number of people those points could serve at that standard. It does not establish how many people a point actually serves, whether its water is safe, or whether households use it, so $R_i$ is not an access measure. Because $k$ scales every LGA equally, it changes the level of $R_i$ but not the ranking. For 2030 we compute $C_i / \text{Pop}_{i,2030}$ with $C_i$ unchanged. This is a population-growth scenario with unchanged supply, not a forecast of water point condition.

\subsection{Record Count and Record Age}
Each in-scope LGA is classed by its number of mapped points: none, 1 to 9, or 10 or more. Ten points is a minimum record-count threshold for ranking. Meeting it does not show that an LGA is adequately mapped, because the completeness of the registry is unknown in every LGA; the threshold only avoids ranking LGAs on a handful of records. Only LGAs that meet the threshold are ranked. LGAs below it are listed separately for verification and mapping, with no capacity imputed. Record age is reported directly rather than as a category: every LGA value is shown with the median year in which its statuses were recorded and the age of that median record in 2025.

\subsection{Empirical Residual Ranges}
For every LGA with mapped points we computed the residual between the observed share of functional points and the share predicted by spatial block cross-validation. Residuals were grouped by the number of mapped points (1 to 9, 10 to 29, 30 to 99 and 100 or more), and the 5th and 95th percentiles of each group were added to the predicted share of every LGA in that group \citep{angelopoulos2023}. The resulting range is carried through to $F_i$ and $R_i$. The residuals come from the same LGAs to which the ranges are applied, so about 90\% of those LGAs fall inside their own range by construction. We therefore call them empirical residual ranges and do not claim validated 90\% coverage. To see how the ranges carry over to LGAs not used to build them, we rebuilt them five times, each time from the residuals of four fifths of the LGAs, and counted how often the observed share of each held-out LGA fell inside its range, first with LGAs assigned to folds at random and then with whole states held out. The ranges describe disagreement between predicted and recorded status among mapped points. They do not include uncertainty in population, in the value of $k$ or in the completeness of the registry, which is likely to be larger.

\subsection{Scope Definition}
WPDx tracks communal and public water points and excludes privately owned household boreholes, which are the dominant source in Lagos and other cities \citep{healy2020}. We therefore restricted the analysis to rural and peri-urban LGAs. An LGA was classified as urban if its 2025 population density was at least 1,000 people per km$^2$. We tested thresholds from 750 to 2,000 per km$^2$ and cross-checked the classification against the density of GRID3 \textit{Built-up Area} blocks.

\subsection{Relative Stress Index}
For ranked LGAs we built a relative stress index from three min-max normalised components \citep{oecd2008}: population growth from 2025 to 2030 ($g$), the predicted share of non-functional points, $1 - \bar{p}_i$ ($f$), and low mapped point density per 1,000 people ($\ell$, inverted and capped at its 95th percentile):
\begin{equation}
S_i = 100 \times (0.35\,\hat{g}_i + 0.40\,\hat{f}_i + 0.25\,\hat{\ell}_i).
\end{equation}
We compared its top 20 LGAs with an equal-weight version. The index orders LGAs relative to one another and has no external validation.

\subsection{Sensitivity Analysis}
We compared the main ranking of $R_i$ with six alternatives, using Spearman rank correlation and the overlap among the 60 lowest LGAs: (1) \textit{Functional, not in use} counted as unavailable, with the model retrained on that target; (2) observed statuses in place of model probabilities; (3) the count of points with $p_j \geq 0.5$; (4) age set to 2025 and normal-year rainfall; (5) raw mapped points per capita; and (6) the kriging estimate from tiles with two or more points, evaluated at the LGA centroid.

\subsection{Field Verification Sample}
Every validation above uses registry records, so none tests the present condition of a point. A small independent inspection would. We drew a stratified random sample of \NVerify{} points from the \NVerifyFrame{} points in \NVerifyLGAs{} priority LGAs (the 60 ranked LGAs with the lowest ratios), stratified by zone (north and south) and predicted probability band (below one third, one third to two thirds, above two thirds), giving \NVerifyStrata{} strata. With this sample the overall functional rate can be estimated to within about \VerifyMargin{} percentage points at 95\% confidence, and calibration can be checked within each probability band. Appendix G sets out the protocol.

\section{4. Experimental Results}

\subsection{Exploratory Analysis}
\begin{figure*}[!t]
    \centering
    \includegraphics[width=0.9\linewidth]{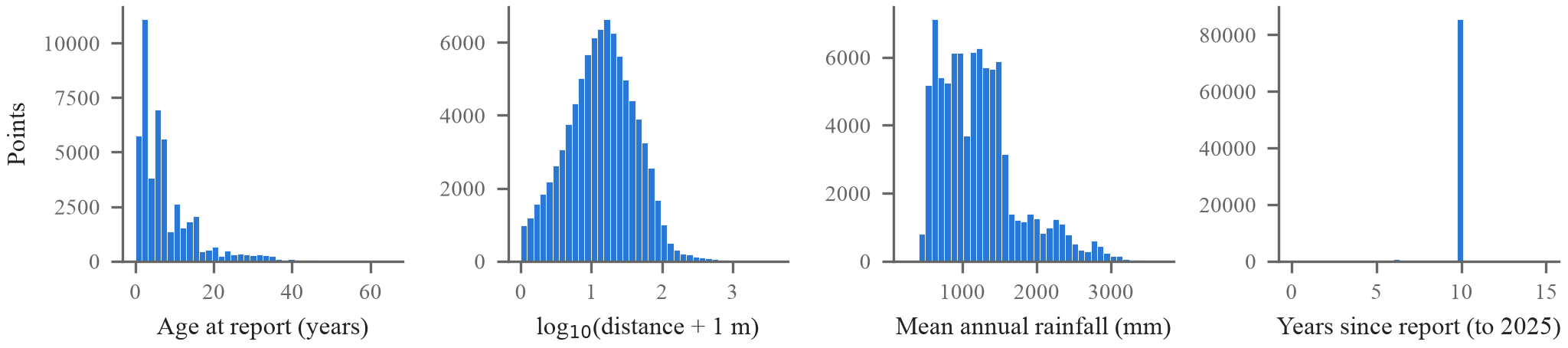}
    \caption{Distributions of age at observation, distance to settlement, long-term rainfall and years since the status was recorded. Years since the status was recorded is ten for almost every point, because \PctRound\% of known statuses date from 2015.}
    \label{fig:dist}
\end{figure*}

\begin{figure}[!htbp]
    \centering
    \includegraphics[width=\linewidth]{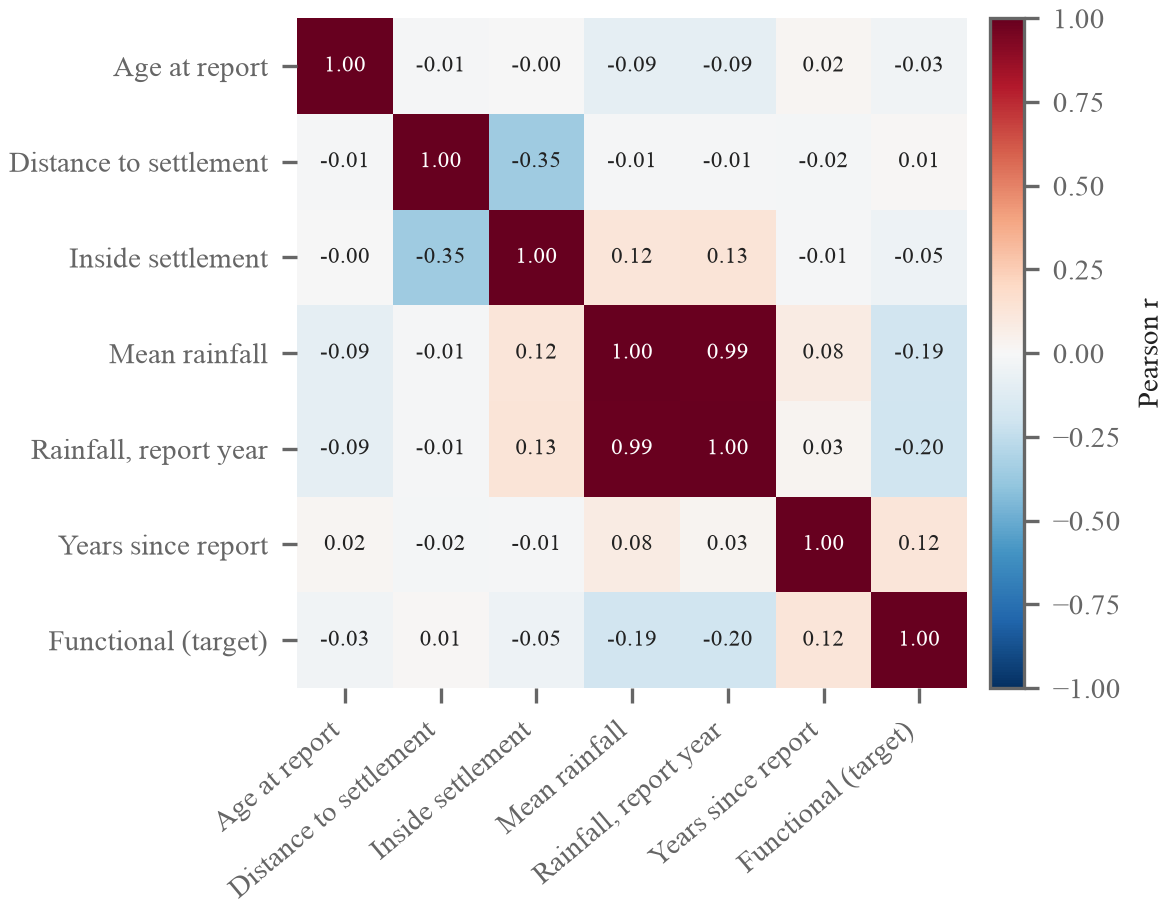}
    \caption{Pearson correlation between the date-aligned features and the functionality target.}
    \label{fig:corr}
\end{figure}

Figure~\ref{fig:dist} shows the distributions of the date-aligned features. Age at observation has a median of \MedAgeObs{} years. The Pearson correlation with the functionality target is \CorrAge{} for age at observation, \CorrDist{} for distance to settlement and \CorrRain{} for long-term rainfall (Figure~\ref{fig:corr}). Years since the report is not a model feature; its correlation with the target mainly reflects which records each survey round contributed.

\subsection{Model Validation}
Table~\ref{tab:val} reports the AUC under each validation scheme. The spatial block AUC was \AUCBlock, against \AUCRandom{} under random cross-validation, and the mean AUC across held-out states was \AUCLoso. Measuring age to 2025 and using 2023 rainfall for every point, instead of aligning both to the report date, gave \AUCOldBlock{} under the same spatial blocks. Adding the nightlight trend to the date-aligned model gave \AUCNight, so the trend adds no predictive value.

\begin{table}[!htbp]
    \centering
    \renewcommand{\arraystretch}{1.35}
    \resizebox{\columnwidth}{!}{%
    \begin{tabular}{|l|c|}
        \hline
        \textbf{Validation scheme} & \textbf{AUC} \\ \hline
        Spatial block CV (primary) & \AUCBlock \\ \hline
        Leave-one-state-out (mean of states) & \AUCLoso \\ \hline
        Random stratified CV (reference) & \AUCRandom \\ \hline
        Age to 2025 and 2023 rainfall, spatial block CV & \AUCOldBlock \\ \hline
        Adding nightlight trend, spatial block CV & \AUCNight \\ \hline
    \end{tabular}
    }
    \caption{Classifier AUC by validation scheme. The temporal hold-out is not listed because its test set contains only non-functional statuses, for which the AUC is undefined.}
    \label{tab:val}
\end{table}

\begin{figure*}[!t]
    \centering
    \includegraphics[width=0.85\linewidth]{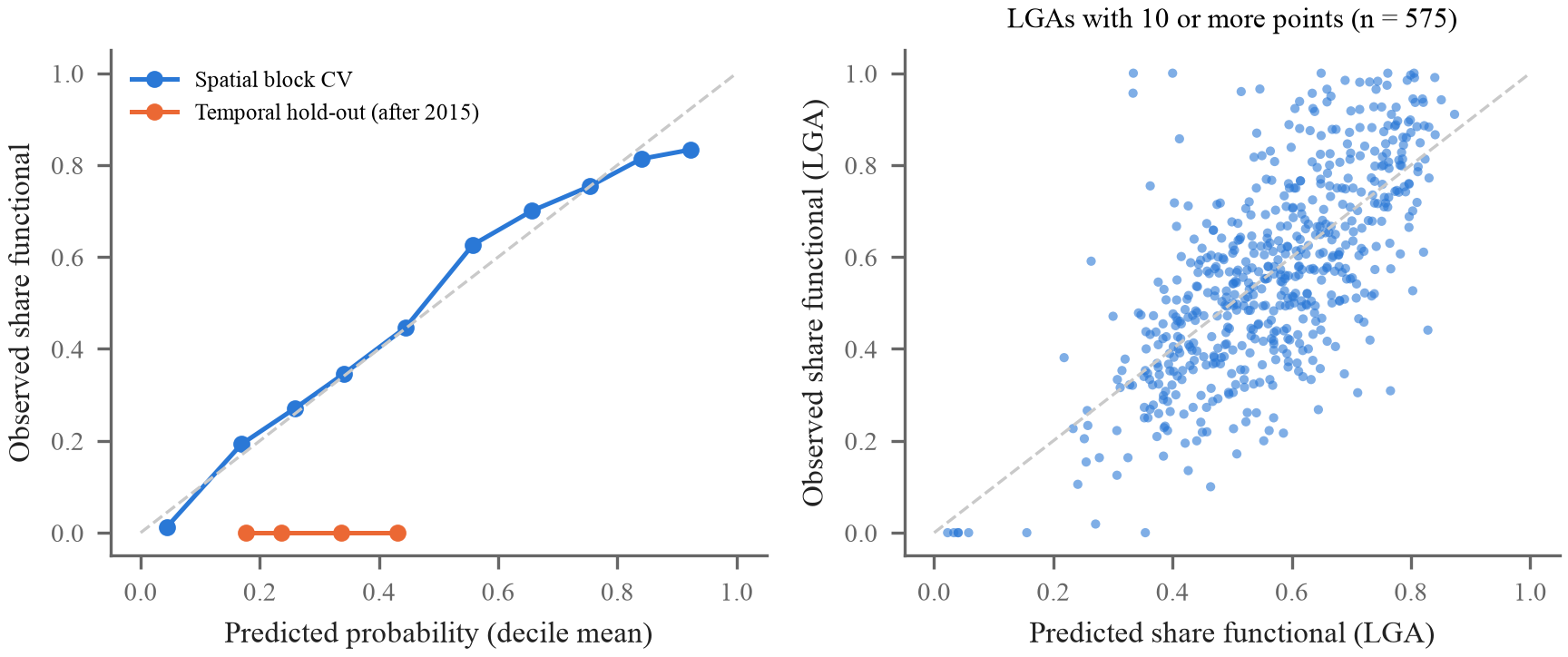}
    \caption{Calibration. Left: observed share functional against predicted probability, by decile, under spatial block cross-validation and for the statuses recorded after 2015 (temporal hold-out, all observed non-functional). Right: predicted against observed share of functional points for LGAs with at least ten points.}
    \label{fig:cal}
\end{figure*}

\textbf{Calibration.} Figure~\ref{fig:cal} compares predicted and observed rates. Under spatial block cross-validation the Brier score was \BrierBlock, the ECE \ECEBlock{} and the calibration slope \CalSlope{}, where a slope below 1 means predictions are too extreme and a slope above 1 means they are too cautious. The model trained on statuses up to \TempCutoff{} gave the \TempNTest{} later statuses a mean predicted probability of \TempMeanP\%, although none was recorded as functional. Later records therefore differ from the training data in a way the features do not capture. At LGA level, the right panel of Figure~\ref{fig:cal} compares the predicted and observed share of functional points for LGAs with at least ten points.

\begin{figure*}[!t]
    \centering
    \includegraphics[width=0.9\linewidth]{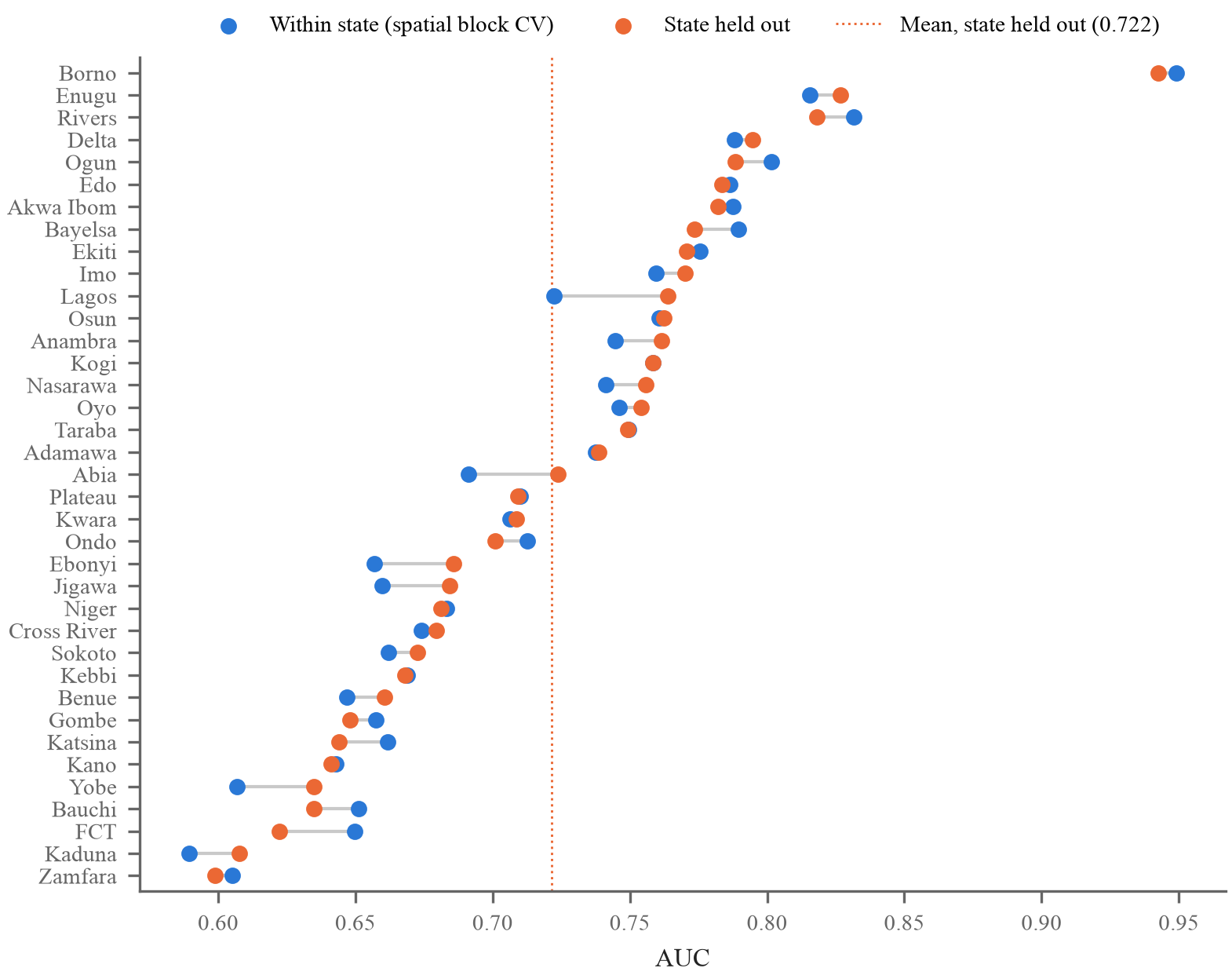}
    \caption{Classifier AUC by state: within state, from spatial block cross-validation, and with the state held out. States are sorted by held-out AUC.}
    \label{fig:stateauc}
\end{figure*}

\textbf{Regional variation.} Leave-one-state-out AUC ranged from \LosoMin{} in \LosoMinState{} to \LosoMax{} in \LosoMaxState{} (Figure~\ref{fig:stateauc}; Appendix C). \ifx\NLosoWeak\ZeroText No state fell below an AUC of 0.70 when held out.\else The \NLosoWeak{} states with a held-out AUC of 0.70 or less were \LosoWeakStates. Estimates in these states should be treated as the least reliable.\fi

\subsection{Aggregation Method Test}
\begin{figure}[!htbp]
    \centering
    \includegraphics[width=\linewidth]{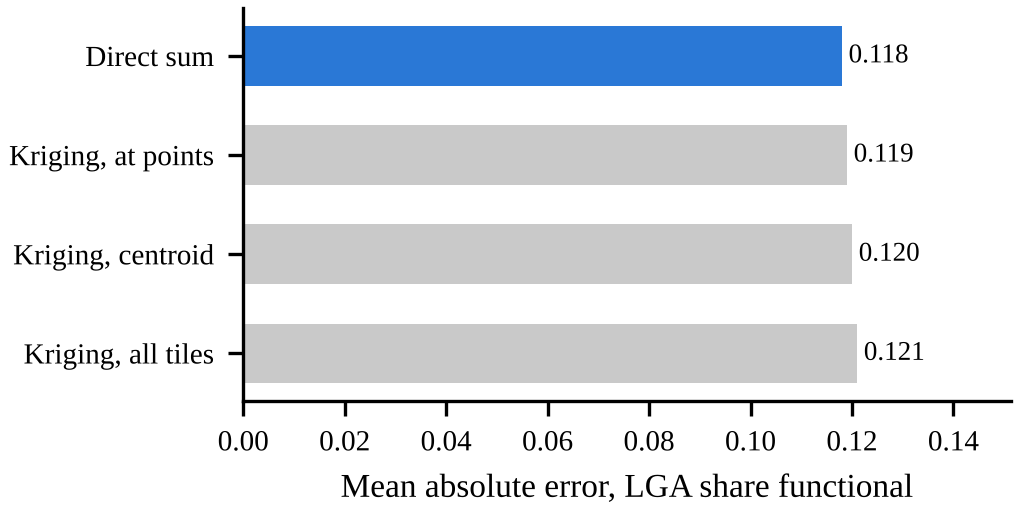}
    \caption{Mean absolute error of the four LGA estimates against observed functionality.}
    \label{fig:agg}
\end{figure}

Table~\ref{tab:agg} compares the four LGA estimates with the observed share of functional points in the \NLGAEval{} LGAs with at least ten points. Of the four, \AggVerdict{} (Figure~\ref{fig:agg}). Dropping single-point tiles removed \PctSingleTiles\% of the \NTilesAll{} tiles, leaving \NTilesTwo, so kriging from those tiles ignores the points in the most sparsely mapped places. Using the centroid assigned each LGA the surface value at one location, which can lie far from where its points are. In the gap-filling test, kriging gave a mean absolute error of \GapMAEKrig{} for LGAs it had not seen, against \GapMAEState{} for the mean of other LGAs in the same state, so kriging \GapVerdict{} a simple state average. We therefore use the direct sum of probabilities and report kriging only as a comparison (fitted variogram: nugget \VarioNugget, partial sill \VarioSill, range \VarioRange$^\circ$).

\begin{table}[!htbp]
    \centering
    \renewcommand{\arraystretch}{1.35}
    \resizebox{\columnwidth}{!}{%
    \begin{tabular}{|l|c|c|c|c|}
        \hline
        \textbf{LGA estimate} & \textbf{MAE} & \textbf{Bias} & \textbf{Spearman} & \textbf{MAE, points} \\ \hline
        Direct mean of point probabilities & \MAEDirect & \BiasDirect & \RhoDirect & \MAECountDirect \\ \hline
        Kriging, 2+ point tiles, centroid & \MAEKOrig & \BiasKOrig & \RhoKOrig & \MAECountKOrig \\ \hline
        Kriging, all tiles, centroid & \MAEKAll & \BiasKAll & \RhoKAll & \MAECountKAll \\ \hline
        Kriging, 2+ point tiles, at the points & \MAEKPts & \BiasKPts & \RhoKPts & \MAECountKPts \\ \hline
    \end{tabular}
    }
    \caption{Error of each LGA estimate against the observed share of functional points (LGAs with at least ten points). MAE and bias are in share units; the last column is the mean absolute error in number of functional points.}
    \label{tab:agg}
\end{table}

\subsection{Population Layer}
After the no-data correction, the national raster totals were 235.1 million for 2025 and 259.9 million for 2030. Summing LGA totals gave 234.8 million and 259.6 million, a difference of about 0.1\% that comes from cells split by LGA boundaries.

\subsection{Scope Definition}
\begin{figure}[!htbp]
    \centering
    \includegraphics[width=\linewidth]{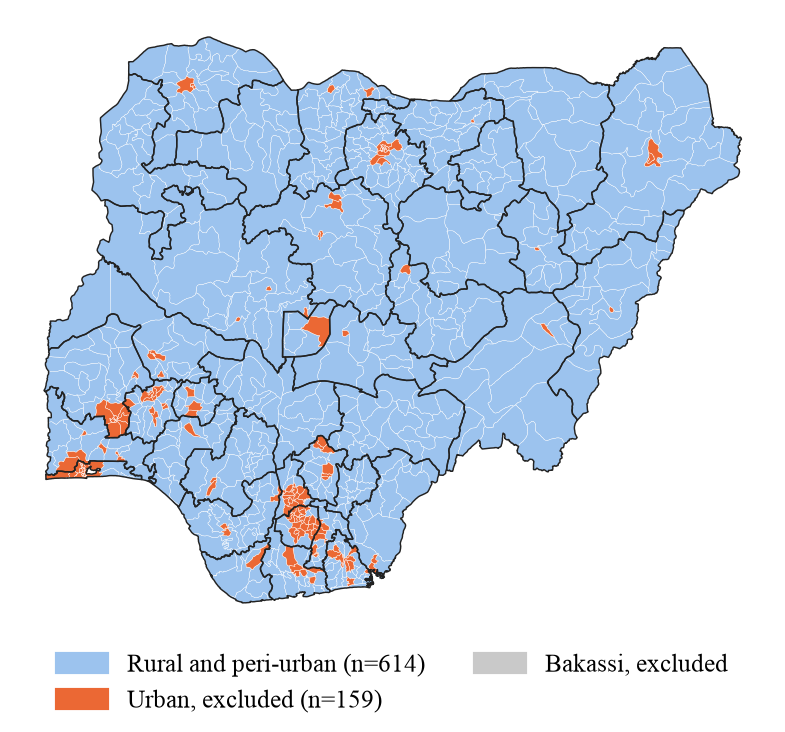}
    \caption{LGAs in scope (rural and peri-urban) and excluded urban LGAs at a threshold of 1,000 people per km$^2$.}
    \label{fig:scope}
\end{figure}

At 1,000 people per km$^2$, 159 LGAs (20.5\%) were classified as urban and 615 as rural or peri-urban. Excluding Bakassi, 614 rural and peri-urban LGAs are in scope. The count changes smoothly across thresholds (Table~\ref{tab:threshold}). Urban LGAs averaged 24.1 GRID3 built-up blocks per km$^2$ and rural or peri-urban LGAs 1.09, a ratio of about 22 to 1 (Figure~\ref{fig:scope}).

\begin{table}[!htbp]
    \centering
    \renewcommand{\arraystretch}{1.3}
    \resizebox{0.40\textwidth}{!}{
    \begin{tabular}{|c|c|c|}
        \hline
        \textbf{Threshold (per km$^2$)} & \textbf{Urban LGAs} & \textbf{Rural/peri-urban} \\ \hline
        750 & 205 (26.5\%) & 569 \\ \hline
        1,000 & 159 (20.5\%) & 615 \\ \hline
        1,250 & 141 (18.2\%) & 633 \\ \hline
        1,500 & 128 (16.5\%) & 646 \\ \hline
        2,000 & 105 (13.6\%) & 669 \\ \hline
    \end{tabular}
    }
    \caption{Sensitivity of the urban classification to the density threshold.}
    \label{tab:threshold}
\end{table}

\subsection{Data-Quality Labels}
\begin{figure}[!htbp]
    \centering
    \includegraphics[width=\linewidth]{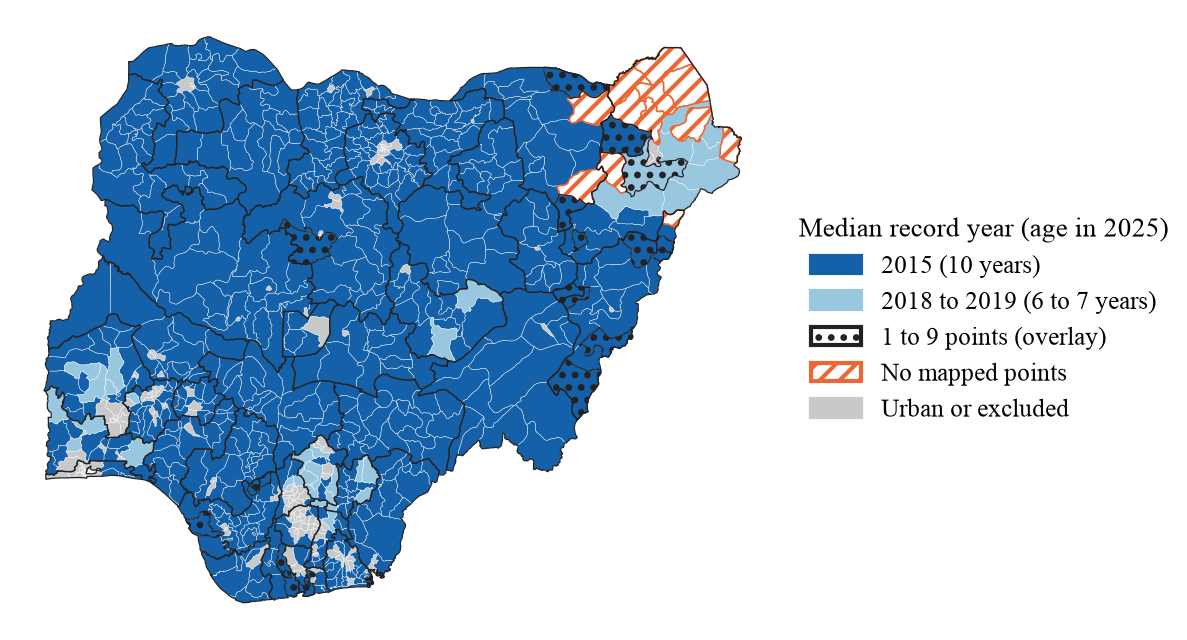}
    \caption{Median year in which the statuses of each LGA's mapped points were recorded, with the age of that record in 2025. LGAs with 1 to 9 points are dotted and LGAs with no mapped points are hatched.}
    \label{fig:dq}
\end{figure}

Of the \NScope{} in-scope LGAs, \NAdequate{} meet the minimum record-count threshold and are ranked, \NSparse{} have 1 to 9 mapped points and \NInsufficient{} have none (Table~\ref{tab:labels}). Meeting the threshold does not show that an LGA's water points are fully mapped. About 86\% of statuses were recorded in 2015 and \PctStalePts\% before it, and in every LGA with mapped points the median record dates from 2015 to 2019, so the typical record was six to ten years old in 2025 (Figure~\ref{fig:dq}). The median record year and its age are listed with every LGA value (Table~\ref{tab:top10}, Appendices D and E).

\begin{table}[!htbp]
    \centering
    \renewcommand{\arraystretch}{1.3}
    \resizebox{\columnwidth}{!}{%
    \begin{tabular}{|l|c|l|}
        \hline
        \textbf{Label} & \textbf{LGAs} & \textbf{Use} \\ \hline
        10 or more points (meets threshold) & \NAdequate & Ranked for inspection \\ \hline
        1 to 9 points & \NSparse & Registry improvement list \\ \hline
        No mapped points & \NInsufficient & Registry improvement list \\ \hline
        Median record year (age in 2025) & 2015 to 2019 (6 to 10 years) & Shown with every value \\ \hline
    \end{tabular}
    }
    \caption{Record count and record age for in-scope LGAs.}
    \label{tab:labels}
\end{table}

\subsection{Nominal Capacity Ranking}
Across the \NRanked{} ranked LGAs, the median nominal capacity ratio $R_i$ was \MedR. At 300 people per point, the expected functional points of the median LGA could nominally serve about \MedRpct\% of its residents. Equivalently, the median share of the population beyond that nominal capacity, $B_i$, was \MedBeyondPct\%, and $B_i$ exceeded 90\% in \PctBeyondNinety\% of ranked LGAs. These two statements describe the same quantity and are consistent. Neither is the share of people without water. The registry records a median of \MedPtsPerK{} mapped points per 1,000 people, against the 2 to 4 per 1,000 implied by the Sphere standard \citep{sphere2018}. A low ratio can reflect incomplete mapping, reliance on sources the registry does not record, such as private boreholes, unimproved sources and surface water, or a real shortage of working points. This analysis cannot distinguish among these explanations, in any LGA. For the same reason, the absolute difference $\text{Pop}_i - C_i$ correlates with population at $r = $ \PopCorr{} and is not used for ranking.

\begin{figure}[!htbp]
    \centering
    \includegraphics[width=\linewidth]{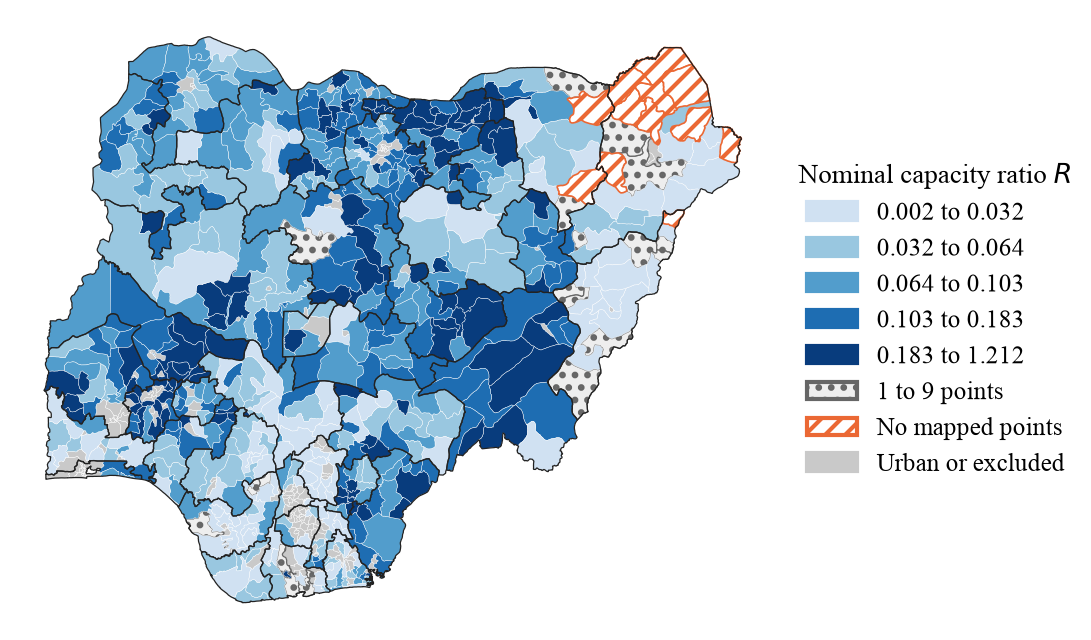}
    \caption{Nominal capacity ratio $R_i$ by LGA (quintiles of ranked LGAs). LGAs with 1 to 9 points are dotted, LGAs with no mapped points are hatched, and urban LGAs and Bakassi are grey.}
    \label{fig:coverage}
\end{figure}

\begin{figure}[H]
    \centering
    \includegraphics[width=0.75\linewidth]{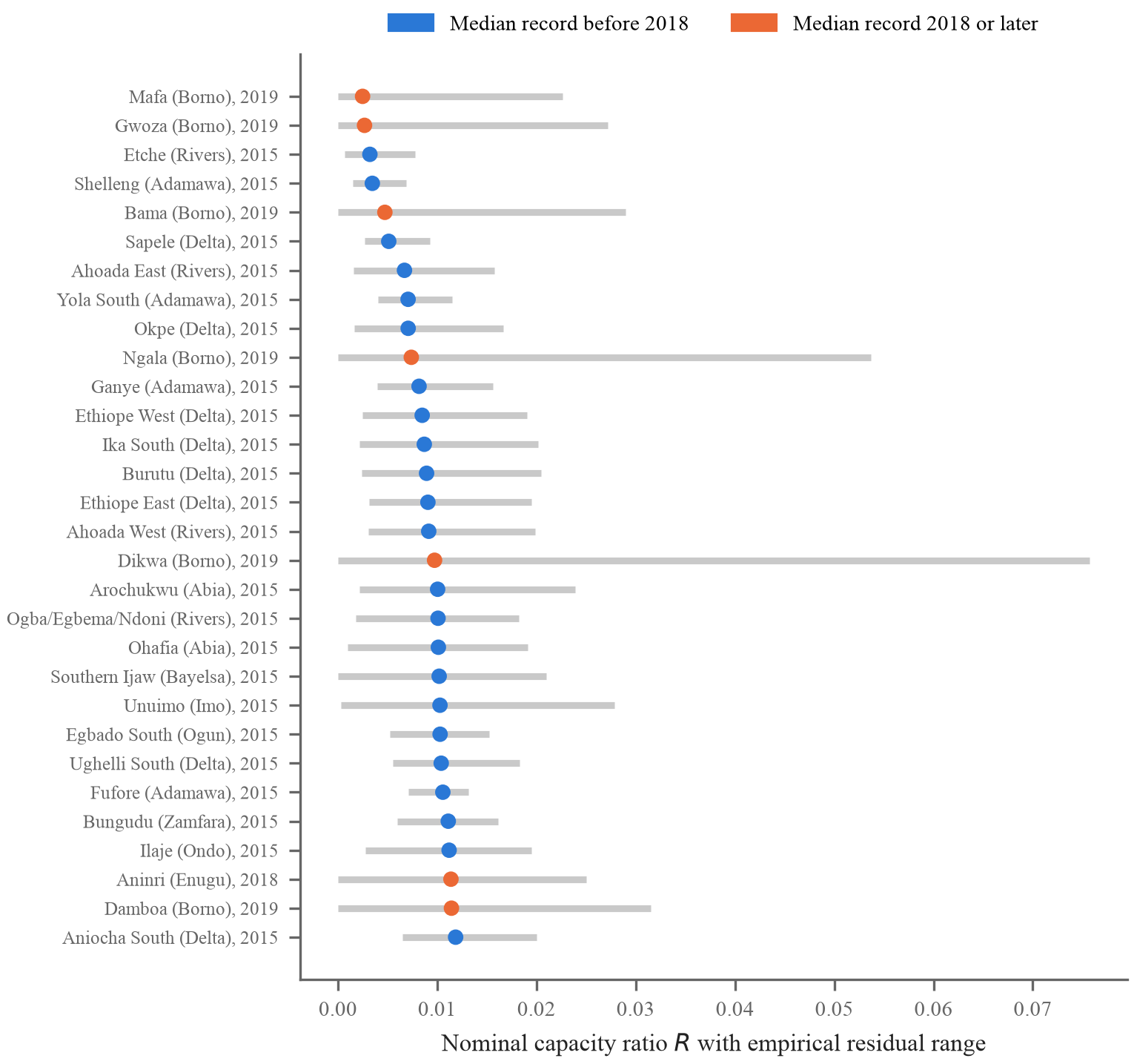}
    \caption{The 30 lowest nominal capacity ratios with empirical residual ranges, coloured by the median year of each LGA's records.}
    \label{fig:top30}
\end{figure}

Figure~\ref{fig:coverage} maps $R_i$ and Table~\ref{tab:top10} lists the ten lowest values with their empirical residual ranges. The median range for the functional share was \MedIntWidthPct{} percentage points wide. When the ranges were rebuilt without the LGAs being checked, \IntCovHeldLGA\% of the \NIntHeld{} held-out LGAs had an observed share inside their range with random folds, and \IntCovHeldState\% with whole states held out (\IntCovHeldLGATen\% and \IntCovHeldStateTen\% for LGAs with ten or more points). This is close to the nominal 90\%, but it is a descriptive check, not a coverage guarantee. Among the 30 lowest LGAs (Appendix D, Figure~\ref{fig:top30}), the most frequent states were \TopStates. For all 30, the upper end of the residual range lies below the median ratio, so their position does not rest on model error alone. Their median records are six to ten years old, and \NBottomLate{} have a median report year of 2018 or 2019, when most records came from the GRID3 survey whose statuses are unknown, so their position rests mainly on predicted rather than recorded statuses. Their low ratios may reflect incomplete mapping, other sources, a real shortage, or the model's predictions for points of unknown status, and only a field check can tell which. The \NRegistry{} LGAs with fewer than ten mapped points are listed in Appendix E for verification and mapping.

\begin{table*}[!t]
    \centering
    \renewcommand{\arraystretch}{1.3}
    \small
    \begin{tabular}{|l|l|r|c|c|c|l|}
        \hline
        \textbf{LGA} & \textbf{State} & \textbf{Pop. 2025} & \textbf{Points} & \textbf{$R_i$} & \textbf{Residual range} & \textbf{Median record year (age)} \\ \hline
        \TabBottomTen
    \end{tabular}
    \caption{The ten ranked LGAs with the lowest nominal capacity ratio. Values describe mapped communal points at the time of their last report and are screening indicators, not access estimates.}
    \label{tab:top10}
\end{table*}

\textbf{2030 scenario.} With supply held at its current estimate, the median ratio falls from \MedR{} to \MedRFuture{} (median change \MedRChange\%). This isolates the effect of population growth and is not a forecast of water point condition.

\subsection{Relative Stress Index}
For the ranked LGAs, the stress index had a mean of \SMean{} and a standard deviation of \SSD, with a range of \SMin{} to \SMax. The highest values were in \STopFive. The top 20 shared \STopOverlap{} LGAs with an equal-weight version.

\subsection{Supply Concentration}
We planned to measure supply concentration with the Herfindahl-Hirschman index (HHI) \citep{rhoades1993}, but WPDx's \texttt{scheme\_id} field is empty for every Nigerian record. A proxy that grouped points sharing a source type within 0.01$^\circ$ cells correlated with the number of mapped points at $\rho = -0.78$, and its normalised form still did at $\rho = -0.47$, so it mostly measured how sparsely an LGA is mapped. Supply concentration could not be assessed, and we report this as a data limitation rather than a finding.

\subsection{Sensitivity of the Ranking}
Table~\ref{tab:sens} summarises the sensitivity checks. The main estimate uses no classification threshold, because it sums probabilities. Counting only points with $p \geq 0.5$ discards the information in the probabilities, and the third row shows how far that thresholded count moves the ranking. Setting age to 2025 changed total expected functional points by \RefShareChange\%. The largest effect comes from the \textit{Functional, not in use} label. Counting the \NNotInUse{} points that carry it as unavailable leaves 82 functional points in the registry, changes total expected functional points by \NotInUseShareChange\%, and gives a ranking that shares only \OvAlt{} of the 60 lowest LGAs with the main one. The ranking therefore depends heavily on how this label is read, and the field verification records whether each sampled point is in use (Appendix G). Coding the \NUnknown{} points of unknown status as non-functional, as WPDx's cleaned label does, gives a ranking that shares \OvUnk{} of the 60 lowest LGAs with the main one.

\begin{table}[!htbp]
    \centering
    \renewcommand{\arraystretch}{1.35}
    \resizebox{\columnwidth}{!}{%
    \begin{tabular}{|l|c|c|}
        \hline
        \textbf{Alternative to the main ranking} & \textbf{Spearman} & \textbf{Bottom-60 overlap} \\ \hline
        \textit{Functional, not in use} as unavailable & \RhoAltSens & \OvAlt{} of 60 \\ \hline
        Observed statuses, no model & \RhoObsSens & \OvObs{} of 60 \\ \hline
        Count of points with $p \geq 0.5$ & \RhoThrSens & \OvThr{} of 60 \\ \hline
        Age set to 2025, normal-year rainfall & \RhoRefSens & \OvRef{} of 60 \\ \hline
        Raw mapped points per capita & \RhoRawSens & \OvRaw{} of 60 \\ \hline
        Kriging, 2+ point tiles, centroid & \RhoKrigSens & \OvKrig{} of 60 \\ \hline
        2015 survey round only (LGAs ranked in both) & \RhoRound & \OvRound{} of 60 \\ \hline
        Unknown statuses coded non-functional, as in WPDx & \RhoUnkSens & \OvUnk{} of 60 \\ \hline
    \end{tabular}
    }
    \caption{Sensitivity of the ranking of $R_i$ to modelling choices.}
    \label{tab:sens}
\end{table}

\subsection{Status Labels and the 2015 Survey Round}
We compared WPDx's cleaned label with its own status field (whether water was present when the point was assessed), the contributor, the report date and the date each record entered WPDx. Two findings follow. First, the post-2015 pattern is largely a coding artefact: \NLateUnknown{} of the \NLate{} records after 2015 come from a GRID3 survey reported in 2018 and 2019 and entered WPDx in 2021, and all have an unknown status that the cleaned label codes as non-functional. We treat them as unknown (Section 3). Second, the \textit{Functional, not in use} label is confined to one contributor, the FMWR inventory of 2015, and covers almost all of its working points. Water was recorded as present at 99.9\% of them, which supports counting them as functional, but the original status text that would show whether they were in use is not in the WPDx export. This label therefore remains unresolved, and we tested how much the results depend on it.

The 2015 round holds \NRound{} statuses (\PctRound\% of points with a known status), of which \RoundPctFunc\% are recorded as functional and \RoundPctNotInUse\% of those as \textit{Functional, not in use}. A model trained and validated on this round alone reached a spatial block AUC of \AUCRoundBlock{} and a mean held-out-state AUC of \AUCRoundLoso{}, against \AUCBlock{} and \AUCLoso{} for all records, with a Brier score of \BrierRound, an ECE of \ECERound{} and a calibration slope of \CalSlopeRound. Ranking the \NRoundRanked{} LGAs with at least ten 2015 points gave a Spearman correlation of \RhoRound{} with the main ranking over the \NRoundCommon{} LGAs ranked in both, and \OvRound{} of the 60 lowest LGAs in common (Table~\ref{tab:sens}). Within the 2015 round, counting \textit{Functional, not in use} points as unavailable left \OvRoundAlt{} of its 60 lowest LGAs in place (Spearman \RhoRoundAlt).

Because the label cannot be resolved from the registry, the rankings depend on the chosen interpretation. The main analysis counts these points as functional. Under the alternative reading, only \OvAlt{} of the 60 priority LGAs remain among the 60 lowest, so any use of the ranking should state which reading it rests on.

\subsection{Household Survey Context}
MICS 2021 measures the share of household members who report using a basic drinking water service in each state in 2021. Our estimates describe the recorded condition of mapped communal points, observed in different years, in rural LGAs. These are different outcomes, populations and periods, so the comparison cannot validate the communal-water estimates. As context only, the state mean of the stress index correlated with MICS basic service at $\rho = $ \MicsRhoS{} ($p = $ \MicsPS, \MicsN{} states), and mean rainfall at $\rho = $ \MicsRhoRain{} ($p = $ \MicsPRain). Appendix F gives the state values.

\section{5. Discussion}
This study set out to build a screening tool for Nigeria's rural water points from an open registry, and to test every step against held-out data. We now recap the main findings, discuss their implications for planning, and set out the limitations and directions for future work.

\subsection{Key Findings}
First, the registry is a historical record, and an uneven one. With a median report year of \RepYearMedian{} and only \PctRecentPts\% of statuses recorded in \RecentCut{} or later, any functionality estimate built on it describes points as they were when last visited. Aligning age and rainfall to the observation date makes the model consistent with what it learns from. Most records after 2015 carry no observed status, and the few that do are all non-functional, so the model could not be tested over time. Nothing in the registry shows present-day condition.

Second, spatial validation sets the realistic level of performance. The spatial block AUC (\AUCBlock) and the mean held-out-state AUC (\AUCLoso) are the figures to quote, and the random cross-validation AUC (\AUCRandom) is shown only for comparison, as Ploton et al. \citep{ploton2020} recommend. \ifx\NLosoWeak\ZeroText No individual state fell below the 0.70 AUC threshold used in Appendix C.\else Performance is weakest in \LosoWeakStates.\fi

Third, the simplest LGA estimate did at least as well as kriging in our tests. Summing calibrated point probabilities needs no threshold and no interpolation, and \AggVerdict{} against observed LGA functionality, although the margin was small (\MAEDirect{} against \MAEKPts{} to \MAEKAll). Kriging's other purpose, filling LGAs without points, is not needed when those LGAs are labelled as insufficient data rather than given a value.

Fourth, the nominal capacity ratio cannot say why it is low. The median ratio of \MedR{} and the median share of \MedBeyondPct\% beyond nominal capacity describe the same quantity. A low value may reflect incomplete mapping, water sources outside the registry or a real shortage of working points, and the analysis cannot distinguish these. The ranking shows where to look, not what will be found there, and it should be read with each LGA's record count and record age.

Fifth, the rankings depend on how one status label is read. Almost every point recorded as functional is labelled \textit{Functional, not in use}. The label comes from one contributor and water was recorded as present at almost all of these points, but whether they were used could not be confirmed. Read literally, the registry holds 82 working points in use; read as a recording convention, it holds \NFunc. The main analysis takes the second reading. Under the first, only \OvAlt{} of the 60 priority LGAs remain among the 60 lowest. Repeating the analysis on the 2015 round alone, which removes every later record, gave a spatial block AUC of \AUCRoundBlock{} and shared \OvRound{} of the 60 lowest LGAs with the main ranking. The field verification sample records whether each point is in use and can settle the question.

\subsection{Implications for Policy Development}
The outputs are best used as a screening tool for inspection and registry improvement, published with each LGA's record count, record age and the label reading used. Inspection teams can start with the 30 lowest-ranked LGAs, all of whose residual ranges lie wholly below the median ratio, and give priority to the \NBottomLate{} whose records date mainly from 2018 or 2019, when most statuses are unknown. A field check there would show whether low mapped capacity reflects incomplete mapping, other sources, a real shortage or the predicted status of points whose status was never recorded. Because the ranking depends on the reading of \textit{Functional, not in use}, LGAs that rank low under both readings are the safest first targets. Registry programmes can start with the \NRegistry{} LGAs with fewer than ten mapped points, where the data cannot support a ranking at all. The verification sample in Appendix G gives a low-cost way to test the estimates in the field before they are used for allocation. The outputs should not be used to state the number of people without water, the number served by any point, or access to safe drinking water. The 2030 scenario shows that population growth alone reduces nominal capacity per person, which supports maintenance and rehabilitation alongside new construction, but it is not a forecast of water point condition.

\subsection{Limitations and Future Work}
Although the pipeline is tested at each step, several limitations remain. The most important is the nature of WPDx: a contributed registry of improved communal points, recorded in different years, that omits private self-supply, unimproved sources and surface water. Its status labels are also uneven: \textit{Functional, not in use} covers almost every functional point, and WPDx's own status field records water as present at almost all of them, but whether they were used could not be confirmed from the registry fields or the label mapping, so the rankings depend on the interpretation chosen. The cleaned label also codes \NUnknown{} records of unknown status as non-functional; we treat them as unknown. The ten-point threshold limits ranking to LGAs with some records, but it does not show that any LGA is fully mapped. Every output describes the WPDx-mapped communal system at the time of its last report.

A second limitation concerns validation. Spatial and state hold-outs test how well the model transfers across places, but all of them use registry records. Because every known status recorded after 2015 is non-functional, the model could not be tested over time, and features tied to the report year, such as rainfall in that year, may partly learn which programme recorded a point rather than its condition. MICS measures a different outcome, population and period, and cannot validate the estimates. Only the proposed field verification can test current condition directly.

A third limitation concerns alignment. Distance to settlement is measured on a recent settlement layer, and the nightlight trend could not be aligned with older observations. Population and the capacity value $k = 300$ carry uncertainty that the residual ranges do not include, and the ranges themselves are empirical, not intervals with validated coverage.

A fourth limitation concerns insecure areas. In parts of the North-East, low or missing mapping is more plausibly a data gap than a measured shortage, and these LGAs appear on the registry improvement list rather than in the ranking.

Future work could address these limits by carrying out the verification sample, by recording re-inspection dates, scheme identifiers and population served in the registry, and by combining the registry with household surveys in a formal small-area estimation framework \citep{rao2015} with explicit uncertainty.

\section{6. Conclusion}
This paper presented a geospatial machine learning pipeline that screens the functionality and nominal capacity of communal water points in Nigeria's rural and peri-urban LGAs from open data. We aligned point age and rainfall to the date each status was recorded, validated the classifier spatially and by state, checked its calibration, and chose a direct sum of probabilities over kriging after testing both against observed functionality. We ranked only LGAs meeting a minimum record-count threshold, attached empirical residual ranges, record counts and record ages to every value, listed LGAs below the threshold separately for verification, and treated 2030 as a population-growth scenario with unchanged supply. The results support a clear framing: outputs built on a contributed, unevenly dated registry are screening indicators that direct inspection and registry improvement. They do not measure people served or access to safe drinking water, and a low value cannot be attributed to incomplete mapping, other sources or a real shortage without a field check. The rankings also depend on how the registry's \textit{Functional, not in use} label is read: under the alternative reading, only \OvAlt{} of the 60 priority LGAs remain. A small field verification of the kind designed here would be the most direct way to test and strengthen them.

\section*{Acknowledgment}
This work is fully funded by Hamoye Foundation, a 501(c)(3) organization.

\onecolumn
\raggedbottom
\section*{Appendix}

\bigskip
This section provides additional details to support the main findings of the paper. The appendices include the feature definitions and their literature basis (Appendix A), the data-quality issues (Appendix B), state-level classifier performance (Appendix C), the 30 ranked LGAs with the lowest nominal capacity ratio (Appendix D), the registry improvement list (Appendix E), the MICS 2021 state values used as context (Appendix F), the field verification protocol (Appendix G), additional figures (Appendix H) and a glossary of acronyms (Appendix I). The code for the full pipeline is available at \url{https://github.com/projects-hamoye-org/AIPITA_NG} for reproducibility.

\section*{A. Feature Definitions}

\begin{table}[H]
\centering
\renewcommand{\arraystretch}{1.5}
\begin{tabular}{|p{1.5in}|p{2.7in}|p{2.1in}|}
\hline
\textbf{Feature} & \textbf{Definition and source} & \textbf{Basis in the literature} \\ \hline
Age at observation & Report year minus the later of installation and rehabilitation year (WPDx); missing values left missing & Functionality declines with installation age \citep{foster2013,foster2020} \\ \hline
Long-term rainfall & Mean annual CHIRPS total over the available years (mm) & Recharge and hydrogeological setting \citep{funk2015} \\ \hline
Rainfall in report year & CHIRPS annual total for the year the status was recorded (mm) & Seasonal and interannual stress on supply \citep{funk2015} \\ \hline
Distance to settlement & Metres to the nearest GRID3 settlement block boundary (UTM 32N); not date-aligned & Proximity shapes use and maintenance \citep{foster2013} \\ \hline
Inside settlement & 1 if the point lies inside a GRID3 settlement block & Same as above \\ \hline
Management type & One-hot encoding of WPDx management categories & Governance and repair arrangements \citep{foster2013,whaley2017} \\ \hline
Nightlight trend (sensitivity only) & Change in VIIRS radiance, Aug. 2018 to Aug. 2023; not alignable with reports before 2018 & Proxy for local economic change \citep{henderson2012,roman2018} \\ \hline
\end{tabular}
\caption{Model features, definitions and literature basis.}
\end{table}

\section*{B. Data-Quality Issues}

Table~\ref{tab:dq} lists data-quality issues found during the study and how each was resolved.

\begin{table}[H]
\centering
\renewcommand{\arraystretch}{1.45}
\begin{tabular}{|p{2.4in}|p{3.9in}|}
\hline
\textbf{Issue} & \textbf{Resolution} \\ \hline
30 points outside all LGA polygons & Assigned with WPDx LGA fields; one spelling (Mai'Adua) reconciled \\ \hline
Coordinate systems differed (EPSG:3857 and EPSG:4326) & GRID3 reprojected to EPSG:4326; distances computed in UTM 32N \\ \hline
87.4\% of points at zero distance to settlement & Distance measured to block boundary; inside-settlement flag added \\ \hline
CHIRPS fill value ($-9999$) & Masked before any calculation \\ \hline
Installation year of 2045; installation after report year (\NNegAge{} records) & Treated as missing for age at observation \\ \hline
Report years spread over \RepYearMin{} to \RepYearMax, about 86\% in 2015 & Features aligned to report year; median record year and age shown for every LGA; validation and ranking repeated on the 2015 round alone \\ \hline
All \NLate{} records after 2015 coded non-functional; \NLateUnknown{} of them (GRID3, 2018 to 2019) have status \textit{Unknown} in WPDx's own field & Unknown statuses excluded from training and validation and given predicted probabilities; WPDx's coding tested as a sensitivity case; the other \TempNTest{} later statuses are all non-functional, so the temporal hold-out is a check only \\ \hline
\textit{Functional, not in use} on \NNotInUse{} of \NFunc{} functional points & All from one contributor (FMWR, 2015); water recorded as present at 99.9\%; counted as functional, alternative reading reported, rankings stated to depend on it \\ \hline
\NConflict{} records whose WPDx status field disagrees with the cleaned label & Cleaned label kept \\ \hline
Cross-validation folds dominated by a few states when records were ordered by state (fold AUC 0.29 to 0.80) & Folds formed by spatial block and by state \\ \hline
WPDx status prediction fields empty & Benchmark against them not possible \\ \hline
WorldPop no-data summed to zero by \texttt{nansum} & No-data restored where no age band held valid data \\ \hline
LGAs with no WPDx points & No imputation; listed for verification \\ \hline
Bakassi: population about 132 & Excluded \\ \hline
36 duplicated rows from joins on LGA names & All joins keyed on \texttt{adm2\_pcode} \\ \hline
Concentration index driven by point count & Not reported; treated as a data limitation \\ \hline
\end{tabular}
\caption{Data-quality issues found during the study and how each was resolved.}
\label{tab:dq}
\end{table}

\section*{C. State-Level Classifier Performance}

\begin{table}[H]
\centering
\renewcommand{\arraystretch}{1.1}
\small
\begin{tabular}{|l|r|c|c|c|}
\hline
\textbf{State} & \textbf{Points} & \textbf{Share functional} & \textbf{AUC within state (spatial block CV)} & \textbf{AUC, state held out} \\ \hline
\TabStateAUC
\end{tabular}
\caption{Classifier performance by state, sorted by held-out AUC. All mapped points are used, including those in urban LGAs.}
\end{table}

\section*{D. The 30 Ranked LGAs with the Lowest Nominal Capacity Ratio}

\begin{table}[H]
\centering
\renewcommand{\arraystretch}{1.1}
\small
\begin{tabular}{|c|l|l|r|c|c|c|l|}
\hline
\textbf{Rank} & \textbf{LGA} & \textbf{State} & \textbf{Pop. 2025} & \textbf{Points} & \textbf{$R_i$} & \textbf{Residual range} & \textbf{Median record year (age)} \\ \hline
\TabBottomThirty
\end{tabular}
\caption{The 30 ranked LGAs (meeting the threshold of ten mapped points) with the lowest nominal capacity ratio, with the median year of their records and its age in 2025. Screening values for inspection, not access estimates.}
\end{table}

\section*{E. Registry Improvement List}

These LGAs have no mapped WPDx points or fewer than ten, so they are not ranked and no capacity is imputed for them. They are the first candidates for mapping and verification.

\begin{table}[H]
\centering
\renewcommand{\arraystretch}{1.1}
\small
\begin{tabular}{|l|l|r|c|c|}
\hline
\textbf{LGA} & \textbf{State} & \textbf{Pop. 2025} & \textbf{Mapped points} & \textbf{Median record year} \\ \hline
\TabInsufficient
\end{tabular}
\caption{In-scope LGAs with fewer than ten mapped points (registry improvement list).}
\end{table}

\section*{F. MICS 2021 Basic Drinking Water Service by State (Context)}

MICS measures household-reported use of a basic drinking water service in 2021. It is a different outcome, population and period from the recorded condition of mapped communal points, so these values are context and not validation.

\begin{table}[H]
\centering
\renewcommand{\arraystretch}{1.12}
\small
\begin{tabular}{|l|c||l|c||l|c||l|c|}
\hline
\textbf{State} & \textbf{\%} & \textbf{State} & \textbf{\%} & \textbf{State} & \textbf{\%} & \textbf{State} & \textbf{\%} \\ \hline
Abia & 81.6 & Delta & 79.1 & Kaduna & 66.0 & Ondo & 79.7 \\ \hline
Adamawa & 56.6 & Ebonyi & 85.7 & Kano & 66.2 & Osun & 91.9 \\ \hline
Akwa Ibom & 72.8 & Edo & 93.5 & Katsina & 66.9 & Oyo & 81.9 \\ \hline
Anambra & 90.9 & Ekiti & 82.7 & Kebbi & 59.0 & Plateau & 50.3 \\ \hline
Bauchi & 55.3 & Enugu & 86.5 & Kogi & 71.4 & Rivers & 85.5 \\ \hline
Bayelsa & 54.2 & Gombe & 51.0 & Kwara & 81.4 & Sokoto & 36.1 \\ \hline
Benue & 51.8 & Imo & 90.8 & Lagos & 97.9 & Taraba & 29.1 \\ \hline
Borno$^\ast$ & 69.2 & Jigawa & 74.6 & Nasarawa & 55.7 & Yobe & 66.1 \\ \hline
Cross River & 51.7 & FCT & 92.9 & Niger & 68.5 & Zamfara & 65.0 \\ \hline
& & & & Ogun & 74.8 & & \\ \hline
\multicolumn{8}{|l|}{$^\ast$MICS surveyed 7 of Borno's 27 LGAs. Source: Table WS.1.2, Nigeria MICS 2021 \citep{mics2021}.} \\ \hline
\end{tabular}
\caption{Percentage of household members using basic drinking water services, by state.}
\end{table}

\section*{G. Field Verification Protocol}

\textbf{Purpose.} To test, on independently and recently inspected points, whether the predicted probabilities match observed functionality and whether the lowest-ranked LGAs reflect a real shortage of functional points or a gap in the registry.

\textbf{Sampling frame and sample.} The frame is the \NVerifyFrame{} mapped points in \NVerifyLGAs{} priority LGAs: the 60 ranked LGAs with the lowest nominal capacity ratio. A stratified random sample of \NVerify{} points was drawn across \NVerifyStrata{} strata defined by zone (north, south) and predicted probability band (low, middle, high), with equal allocation so that calibration can be checked within each band. The sample is saved as \texttt{verification\_sample.csv} with coordinates, the last report date and the predicted probability. Points in LGAs on the registry improvement list (Appendix E) should be added to a separate mapping exercise.

\textbf{Data to record.} For each point: date of visit; whether the point exists at the mapped location; whether it delivers water on the day of visit; the reason if it does not; whether it is in use; the approximate number of households using it; technology and management; and the GPS position and a photograph. Unmapped functional points within 500 m of a sampled point should also be recorded, as evidence of registry completeness.

\textbf{Analysis.} Compare the observed functional rate with the mean predicted probability in each stratum; compute the Brier score and calibration slope on the sample; and, for priority LGAs with enough sampled points, check whether the observed rate falls inside the LGA's empirical residual range. With \NVerify{} points the overall functional rate is estimated to within about \VerifyMargin{} percentage points at 95\% confidence.

\section*{H. Additional Figures}

\begin{figure}[H]
    \centering
    \includegraphics[width=0.75\linewidth]{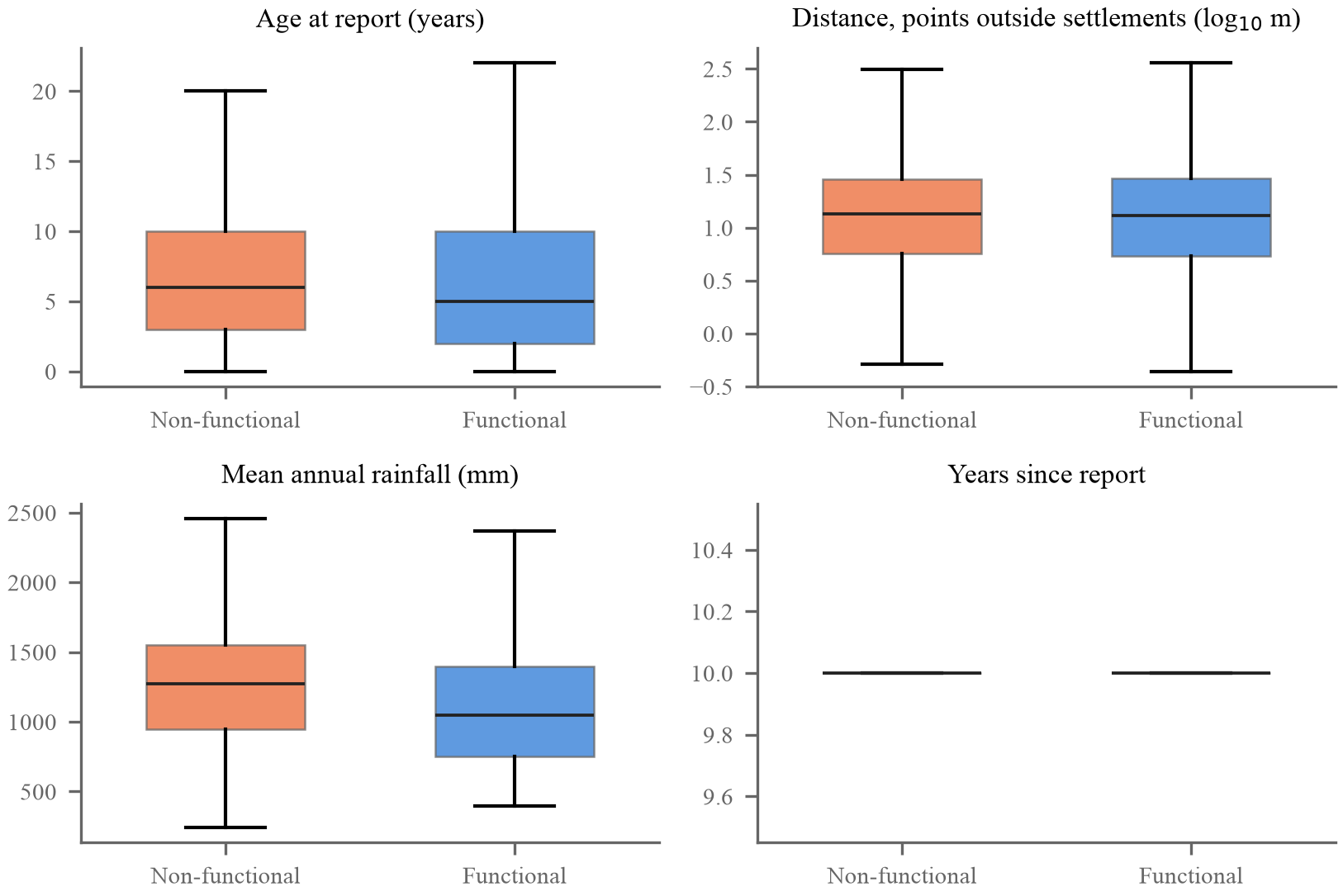}
    \caption{Distribution of the date-aligned features by functionality status (outliers hidden). Years since the status was recorded is ten in both groups, because \PctRound\% of known statuses date from 2015, so it does not separate them.}
\end{figure}

\begin{figure}[H]
    \centering
    \includegraphics[width=0.65\linewidth]{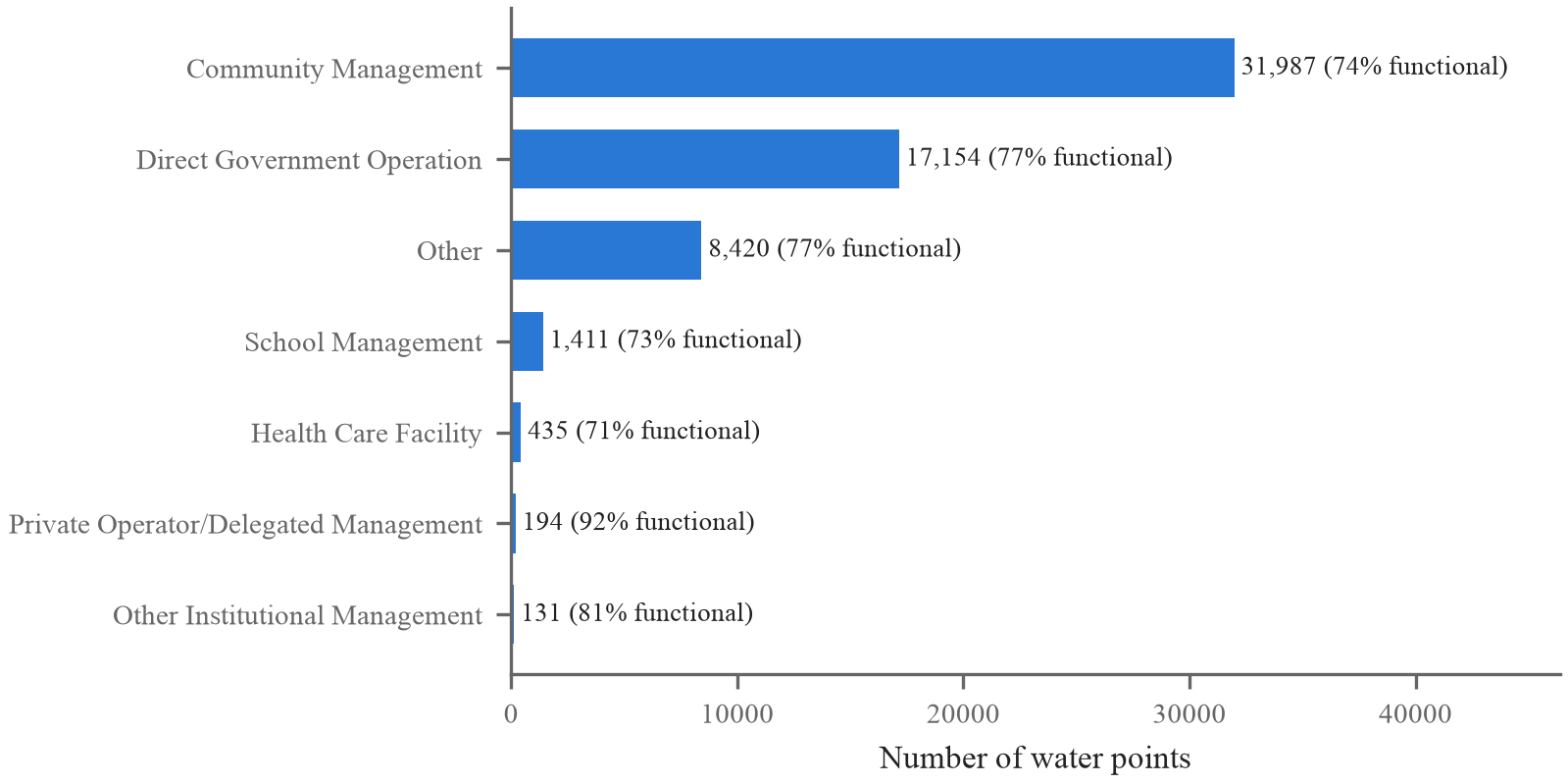}
    \caption{Water points with a recorded management type (59,732 of 99,245), with the share recorded as functional. The remaining 39,513 points have no management type.}
\end{figure}

\begin{figure}[H]
    \centering
    \includegraphics[width=0.8\linewidth]{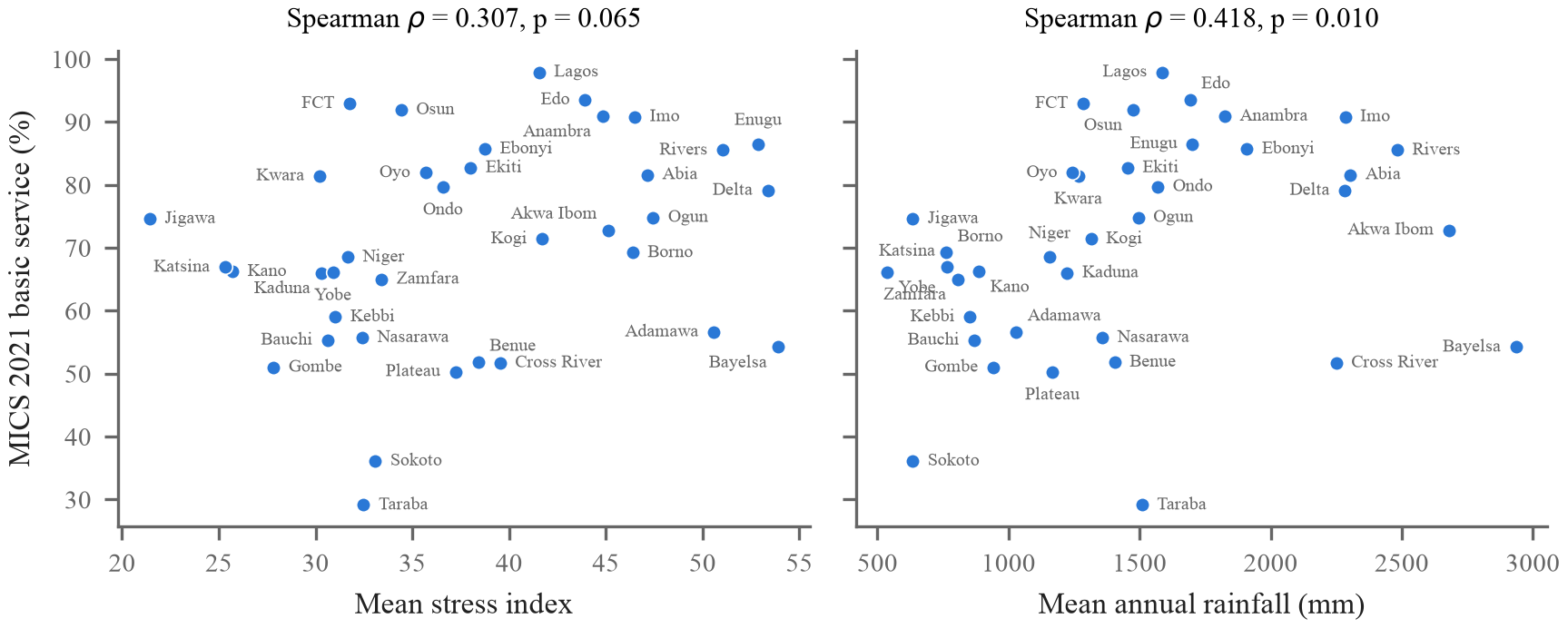}
    \caption{State-level context: MICS 2021 basic drinking water service against the mean stress index (left) and mean rainfall (right). Not a validation.}
\end{figure}

\clearpage
\section*{I. Glossary of Acronyms}

\begin{table}[H]
\centering
{\renewcommand{\arraystretch}{1.5}
\begin{tabular}{ll}
\hline
\textbf{Acronyms} & \textbf{Description} \\
\hline
AUC & Area Under the Receiver Operating Characteristic Curve \\
CHIRPS & Climate Hazards Group InfraRed Precipitation with Station data \\
COD-AB & Common Operational Dataset, Administrative Boundaries \\
CV & Cross-Validation \\
ECE & Expected Calibration Error \\
FCT & Federal Capital Territory \\
GRID3 & Geo-Referenced Infrastructure and Demographic Data for Development \\
HHI & Herfindahl-Hirschman Index \\
JMP & WHO/UNICEF Joint Monitoring Programme \\
LGA & Local Government Area \\
MAE & Mean Absolute Error \\
MICS & Multiple Indicator Cluster Survey \\
OCHA & UN Office for the Coordination of Humanitarian Affairs \\
UTM & Universal Transverse Mercator \\
VIIRS & Visible Infrared Imaging Radiometer Suite \\
WASH & Water, Sanitation and Hygiene \\
WPDx & Water Point Data Exchange \\
\hline
\end{tabular}
}
\caption{Acronyms and their full descriptions.}
\end{table}


\begin{thebibliography}{99}

\bibitem{un2015}
United Nations General Assembly, “Transforming our world: The 2030 agenda for sustainable development,” United Nations, New York, NY, USA, Resolution A/RES/70/1, Oct. 2015. [Online]. Available: https://sdgs.un.org/2030agenda

\bibitem{jmp2023}
WHO and UNICEF, \emph{Progress on Household Drinking Water, Sanitation and Hygiene 2000-2022: Special Focus on Gender}. Geneva, Switzerland: World Health Organization and United Nations Children's Fund, 2023. [Online]. Available: https://washdata.org

\bibitem{mics2021}
National Bureau of Statistics and United Nations Children's Fund, “2021 Multiple Indicator Cluster Survey/National Immunization Coverage Survey (MICS-NICS) report,” NBS and UNICEF, Abuja, Nigeria, Aug. 2022. [Online]. Available: https://mics.unicef.org

\bibitem{foster2020}
T. Foster, S. Furey, B. Banks, and J. Willetts, “Functionality of handpump water supplies: A review of data from sub-Saharan Africa and the Asia-Pacific region,” \emph{Int. J. Water Resour. Dev.}, vol. 36, no. 5, pp. 855-869, 2020, doi:10.1080/07900627.2018.1543117.

\bibitem{foster2013}
T. Foster, “Predictors of sustainability for community-managed handpumps in sub-Saharan Africa: Evidence from Liberia, Sierra Leone, and Uganda,” \emph{Environ. Sci. Technol.}, vol. 47, no. 21, pp. 12037-12046, Nov. 2013, doi:10.1021/es402086n.

\bibitem{whaley2017}
L. Whaley and F. Cleaver, “Can ‘functionality’ save the community management model of rural water supply?” \emph{Water Resour. Rural Dev.}, vol. 9, pp. 56-66, Jun. 2017, doi:10.1016/j.wrr.2017.04.001.

\bibitem{wpdx2024}
Global Water Challenge, “Water Point Data Exchange (WPdx-Plus): Global water point data,” 2024. [Online]. Available: https://www.waterpointdata.org

\bibitem{tatem2017}
A. J. Tatem, “WorldPop, open data for spatial demography,” \emph{Sci. Data}, vol. 4, Art. no. 170004, Jan. 2017, doi:10.1038/sdata.2017.4.

\bibitem{bondarenko2025}
M. Bondarenko \emph{et al.}, “Estimates of 2015-2030 total number of people per grid square broken down by gender and age groupings at a resolution of 30 arc (approximately 1km at the equator), R2025A version v1,” WorldPop, Univ. of Southampton, Southampton, U.K., 2025, doi:10.5258/SOTON/WP00842.

\bibitem{funk2015}
C. Funk \emph{et al.}, “The climate hazards infrared precipitation with stations: A new environmental record for monitoring extremes,” \emph{Sci. Data}, vol. 2, Art. no. 150066, Dec. 2015, doi:10.1038/sdata.2015.66.

\bibitem{roman2018}
M. O. Rom\'{a}n \emph{et al.}, “NASA's Black Marble nighttime lights product suite,” \emph{Remote Sens. Environ.}, vol. 210, pp. 113-143, Jun. 2018, doi:10.1016/j.rse.2018.03.017.

\bibitem{healy2020}
A. Healy \emph{et al.}, “Domestic groundwater abstraction in Lagos, Nigeria: A disjuncture in the science-policy-practice interface?” \emph{Environ. Res. Lett.}, vol. 15, no. 4, Art. no. 045006, Apr. 2020, doi:10.1088/1748-9326/ab7463.

\bibitem{gronwall2020}
J. Gr\"{o}nwall and K. Danert, “Regarding groundwater and drinking water access through a human rights lens: Self-supply as a norm,” \emph{Water}, vol. 12, no. 2, Art. no. 419, Feb. 2020, doi:10.3390/w12020419.

\bibitem{hamoye2026aipita}
Hamoye Foundation, “Age-banded population forecasts and cross-sector capacity gaps in Nigeria, 2025-2030 (education, employment, energy, food, transport, core infrastructure),” Program plan, Hamoye Foundation, 2026.

\bibitem{openhamoye2026}
OpenHamoye Research Team, “Infrastructure stress at national scale: A multi-resolution composite index for Nigeria's 762,751 kilometre-scale grid cells,” OpenHamoye, Tech. Rep., Apr. 2026.

\bibitem{lbd2020}
Local Burden of Disease WaSH Collaborators, “Mapping geographical inequalities in access to drinking water and sanitation facilities in low-income and middle-income countries, 2000-17,” \emph{Lancet Glob. Health}, vol. 8, no. 9, pp. e1162-e1185, Sep. 2020, doi:10.1016/S2214-109X(20)30278-3.

\bibitem{wilson2017}
D. L. Wilson, J. R. Coyle, and E. A. Thomas, “Ensemble machine learning and forecasting can achieve 99\% uptime for rural handpumps,” \emph{PLoS ONE}, vol. 12, no. 11, Art. no. e0188808, Nov. 2017, doi:10.1371/journal.pone.0188808.

\bibitem{chen2016xgboost}
T. Chen and C. Guestrin, “XGBoost: A scalable tree boosting system,” in \emph{Proc. 22nd ACM SIGKDD Int. Conf. Knowl. Discov. Data Min.}, San Francisco, CA, USA, Aug. 2016, pp. 785-794, doi:10.1145/2939672.2939785.

\bibitem{roberts2017}
D. R. Roberts \emph{et al.}, “Cross-validation strategies for data with temporal, spatial, hierarchical, or phylogenetic structure,” \emph{Ecography}, vol. 40, no. 8, pp. 913-929, Aug. 2017, doi:10.1111/ecog.02881.

\bibitem{ploton2020}
P. Ploton \emph{et al.}, “Spatial validation reveals poor predictive performance of large-scale ecological mapping models,” \emph{Nat. Commun.}, vol. 11, Art. no. 4540, Sep. 2020, doi:10.1038/s41467-020-18321-y.

\bibitem{leasure2020}
D. R. Leasure, W. C. Jochem, E. M. Weber, V. Seaman, and A. J. Tatem, “National population mapping from sparse survey data: A hierarchical Bayesian modeling framework to account for uncertainty,” \emph{Proc. Natl. Acad. Sci. USA}, vol. 117, no. 39, pp. 24173-24179, Sep. 2020, doi:10.1073/pnas.1913050117.

\bibitem{henderson2012}
J. V. Henderson, A. Storeygard, and D. N. Weil, “Measuring economic growth from outer space,” \emph{Am. Econ. Rev.}, vol. 102, no. 2, pp. 994-1028, Apr. 2012, doi:10.1257/aer.102.2.994.

\bibitem{elvidge2009}
C. D. Elvidge \emph{et al.}, “A fifteen year record of global natural gas flaring derived from satellite data,” \emph{Energies}, vol. 2, no. 3, pp. 595-622, Aug. 2009, doi:10.3390/en20300595.

\bibitem{jean2016}
N. Jean, M. Burke, M. Xie, W. M. Davis, D. B. Lobell, and S. Ermon, “Combining satellite imagery and machine learning to predict poverty,” \emph{Science}, vol. 353, no. 6301, pp. 790-794, Aug. 2016, doi:10.1126/science.aaf7894.

\bibitem{oliver1990}
M. A. Oliver and R. Webster, “Kriging: A method of interpolation for geographical information systems,” \emph{Int. J. Geogr. Inf. Syst.}, vol. 4, no. 3, pp. 313-332, 1990, doi:10.1080/02693799008941549.

\bibitem{rao2015}
J. N. K. Rao and I. Molina, \emph{Small Area Estimation}, 2nd ed. Hoboken, NJ, USA: Wiley, 2015.

\bibitem{oecd2008}
OECD and European Commission Joint Research Centre, \emph{Handbook on Constructing Composite Indicators: Methodology and User Guide}. Paris, France: OECD Publishing, 2008, doi:10.1787/9789264043466-en.

\bibitem{cattaneo2021}
A. Cattaneo, A. Nelson, and T. McMenomy, “Global mapping of urban-rural catchment areas reveals unequal access to services,” \emph{Proc. Natl. Acad. Sci. USA}, vol. 118, no. 2, Art. no. e2011990118, Jan. 2021, doi:10.1073/pnas.2011990118.

\bibitem{weiss2018}
D. J. Weiss \emph{et al.}, “A global map of travel time to cities to assess inequalities in accessibility in 2015,” \emph{Nature}, vol. 553, no. 7688, pp. 333-336, Jan. 2018, doi:10.1038/nature25181.

\bibitem{grid3settlement}
GRID3, “GRID3 NGA - Settlement Extents v4.1,” Center for International Earth Science Information Network (CIESIN), Columbia Univ., New York, NY, USA, 2026. [Online]. Available: https://data.grid3.org

\bibitem{ochacodab}
UN Office for the Coordination of Humanitarian Affairs, “Nigeria - Subnational administrative boundaries (COD-AB),” Humanitarian Data Exchange, 2026. [Online]. Available: https://data.humdata.org/dataset/cod-ab-nga

\bibitem{icj2002}
International Court of Justice, “Land and maritime boundary between Cameroon and Nigeria (Cameroon v. Nigeria: Equatorial Guinea intervening), Judgment,” \emph{I.C.J. Reports}, pp. 303-458, Oct. 2002.

\bibitem{pedregosa2011}
F. Pedregosa \emph{et al.}, “Scikit-learn: Machine learning in Python,” \emph{J. Mach. Learn. Res.}, vol. 12, pp. 2825-2830, Oct. 2011. [Online]. Available: https://jmlr.org/papers/v12/pedregosa11a.html

\bibitem{sphere2018}
Sphere Association, \emph{The Sphere Handbook: Humanitarian Charter and Minimum Standards in Humanitarian Response}, 4th ed. Geneva, Switzerland: Sphere Association, 2018. [Online]. Available: https://spherestandards.org/handbook

\bibitem{rhoades1993}
S. A. Rhoades, “The Herfindahl-Hirschman index,” \emph{Fed. Reserve Bull.}, vol. 79, no. 3, pp. 188-189, Mar. 1993.


\bibitem{niculescu2005}
A. Niculescu-Mizil and R. Caruana, “Predicting good probabilities with supervised learning,” in \emph{Proc. 22nd Int. Conf. Mach. Learn. (ICML)}, Bonn, Germany, Aug. 2005, pp. 625-632, doi:10.1145/1102351.1102430.

\bibitem{brier1950}
G. W. Brier, “Verification of forecasts expressed in terms of probability,” \emph{Mon. Weather Rev.}, vol. 78, no. 1, pp. 1-3, Jan. 1950, doi:10.1175/1520-0493(1950)078\textless{}0001:VOFEIT\textgreater{}2.0.CO;2.

\bibitem{angelopoulos2023}
A. N. Angelopoulos and S. Bates, “Conformal prediction: A gentle introduction,” \emph{Found. Trends Mach. Learn.}, vol. 16, no. 4, pp. 494-591, Mar. 2023, doi:10.1561/2200000101.

\end{thebibliography}
\end{document}